\documentclass[10pt,twocolumn]{article}
\usepackage[top=2cm, bottom=2cm, left=2cm, right=1.2cm]{geometry}
\usepackage{times}  %
\usepackage[sort&compress,numbers]{natbib}  %
\usepackage[hyphens]{url}
\usepackage{graphicx}  %
\usepackage[keeplastbox]{flushend}
\usepackage[hyphens]{url}  %
\usepackage{graphicx} %
\usepackage{tikzsymbols}
\usepackage{booktabs} 
\usepackage[small,bf]{caption} 

\usepackage{multirow}
\usepackage[flushleft]{threeparttable}
\usepackage{setspace}

\usepackage{csquotes}

\AtBeginDocument{%
  \providecommand\BibTeX{{%
    \normalfont B\kern-0.5em{\scshape i\kern-0.25em b}\kern-0.8em\TeX}}}

\newcommand{\descr}[1]{\smallskip\noindent\textbf{\em #1}}

\let\oldbibliography\thebibliography
\renewcommand{\thebibliography}[1]{%
  \oldbibliography{#1}%
  \setlength{\itemsep}{2pt}%
}

\usepackage[hang,flushmargin]{footmisc}
\renewcommand{\footnotesize}{\fontsize{8}{9}\selectfont}
\usepackage[compact]{titlesec}
\titlespacing*{\section}{0pt}{*2.5}{2.5pt}
\titlespacing*{\subsection}{0pt}{*2}{2pt}
\usepackage{xspace}

\makeatletter
\def\url@leostyle{%
  \@ifundefined{selectfont}{\def\UrlFont{}}%
  {\def\UrlFont{}}%
}
\makeatother
\usepackage[hyphenbreaks]{breakurl}

\usepackage[bookmarks=true, bookmarksnumbered=true, colorlinks=true, linkcolor=linkcol, citecolor=citecol, urlcolor=urlcol, hypertexnames=true]{hyperref}

\definecolor{darkgreen}{RGB}{0, 100, 0}
\definecolor{linkcol}{rgb}{0.3,0,0}
\definecolor{citecol}{rgb}{0.3,0,0}
\definecolor{urlcol}{rgb}{0.3,0,0}

\makeatletter
\def\url@leostyle{%
  \@ifundefined{selectfont}{\def\UrlFont{\small}}%
  {\def\UrlFont{}}%
}
\makeatother
\begin{document}

\title{\LARGE \bf ``I'm a Professor, which isn't usually a dangerous job'': Internet-Facilitated Harassment and its Impact on Researchers}

\author{Periwinkle Doerfler$^1$, Andrea Forte$^2$, Emiliano De Cristofaro$^3$,\\
Gianluca Stringhini$^4$, Jeremy Blackburn$^5$, and Damon McCoy$^1$\\[0.5ex]
\normalsize $^1$New York University Tandon School of Engineering,
$^2$Drexel University,
$^3$UCL \& Alan Turing Institute,\\[-0.5ex]
\normalsize $^4$Boston University,
$^5$Binghamton University
}

\date{}

\maketitle

\begin{abstract}

While the Internet has dramatically increased the exposure that research can receive, it has also facilitated harassment against scholars.
To understand the impact that these attacks can have on the work of researchers, we perform a series of systematic interviews with researchers including academics, journalists, and activists, who have experienced targeted, Internet-facilitated harassment. 
We provide a framework for understanding the types of harassers that target researchers, the harassment that ensues, and the personal and professional impact on individuals and academic freedom.
We then study preventative and remedial strategies available, and the institutions that prevent some of these strategies from being more effective. 
Finally, we discuss the ethical structures that could facilitate more equitable access to participating in research without serious personal suffering.

\end{abstract}

\section{Introduction}

Impactful research and ideas which upend social orders and status quo have often provoked backlash. Physical violence, or the threat thereof, has been used as a tool to silence academics for centuries; consider the untimely demise of Archimedes or Galileo. 
While Western researchers have largely avoided these issues in recent history\footnote{With notable exceptions, including the McCarthy Era.}, there are new, more pervasive threats posed to research communities. 
From terrorists to trolls, the free flow of (mis)information on the Internet has changed the landscape of dangers facing researchers. 
In fact, the authors of this paper have experienced harassment as a result of our research, and know many others who have as well. %

Online harassment is well studied in the context of cyber-bullying~\cite{ioannou2018risk} and gaming~\cite{blackburn2014stfu}, as are individual harassment communities~\cite{flores2018mobilizing,kekscucks} and the personal impacts of online abuse~\cite{quteprints1925}. 
Some resources exist for supporting researchers experiencing online harassment~\cite{DSguide} as do essays exploring the issues~\cite{doi:10.1177/2056305118768302,Philipp2020}; however, to the best of our knowledge, no empirically grounded research examines the ways Internet-facilitated harassment is impacting research communities. 

In addition to the personal repercussions that people face as a result of harassment, researchers face additional professional constraints that prevent them from taking certain mitigation steps, such as reducing their visibility. 
Especially for early-career academics, visibility and publication are required for impactful research and to achieve professional goals. 
For senior faculty, these needs may not be as dire, but reduced visibility and self-censorship impact the types of research the academy produces. 
As Internet-facilitated harassment of researchers is a phenomenon that appears to be growing, it is critical to understand who it is happening to, when, how, and why, to avoid researchers being blindsided by it, and to be able to provide recommendations that balance the need for the personal security of individuals experiencing harassment with the principles of academic freedom. 
One of our interview participants summarized this succinctly:

\begin{displayquote}
``It seems like this is happening more and more to folks and they have no resources to figure it out. I think we should be talking about it before it happens because if you are being public or doing work that might be controversial, you are going to risk harassment.'' --- \textit{Gender and Women's Studies Professor}
\end{displayquote}

Internet-facilitated harassment takes many forms. During this project, a teacher in France was beheaded for conducting a lesson on religious freedom in which he showed a controversial cartoon. 
The assailant traveled from another part of the country, having seen an online campaign by parents of students in the class, pushing for the teacher to be fired~\cite{CNNFranceTeacher}. 
``Cancel Culture'' is one of the greatest buzzwords of this era -- shorthand for what some see as righteous outings and others view as Twitter lynchings~\cite{harpers_letter}. 
These disparate examples underscore how Internet-facilitated harassment is an emerging, changing phenomenon, and one that is increasingly occurring in, and being discussed in, the public sphere. 
Old forms of harassment resurface at a new scale through online organization, while other threats are novel. 
We encountered researchers whose careers predate the Internet and who expressed that the occasional ``crank letter'' is not unexpected, but being doxed~\cite{doxingConceptual} and receiving thousands of them at one's home is a new phenomenon. 
A politician calling for a controversial professor to be fired is unsurprising; dozens of phone calls directed to an academic department demanding a firing, spurred by online pseudo-journalism and rapidly disseminated on social media, however, may be more difficult to contend with.

\descr{Roadmap.} We begin by asking how Internet-facilitated harassment affects researchers and others whose work involves publicizing knowledge and ideas. 
What kind of harassment do they experience, what is its effect on their work, and what strategies do they use to cope with these experiences and threats? 
To address these questions, we conduct a series of interviews with 17 students, faculty, administrators, journalists, and activists. 

In our analysis, we examine the emerging trend of Internet-facilitated harassment of researchers and other public figures. 
We discuss the types of work that led to harassment, its nature and scope, and the harassers' motivations.
We analyze personal and professional ramifications of harassment, particular vulnerabilities in the research world, and the ways in which common resolution strategies are not tenable in academia. Then, we discuss the implications of harassment for knowledge creation and academic freedom, and the compounding effect of existing marginalization and the structure of tenure. Finally, we offer recommendations and suggestions for future work.

\descr{Main Contributions.}
In this paper, we provide a multi-faceted analysis of the problems arising from the harassment of researchers and public figures.
In particular, we make the following contributions:
\begin{itemize}
\item We introduce a motivation-driven model of harassers that enables us to understand and classify the types of harassment that researchers are experiencing.
\item We examine the negative effects of harassment on people's careers, research agendas, and academic freedom, with a particular focus on the role played by institutions.
\item We offer a set of recommendations for individual PIs, institutions, and other organizations for anticipating, preventing, and responding to harassment.
\end{itemize}

\section{Background and Related Work}

In this section, we provide a review of the different kinds of Internet-facilitated harassment as well as countermeasures, with a particular focus on the experiences of researchers.

\subsection{Online Harassment}
Motivated by the increasing relevance of safety issues on the Web, diverse research communities have contributed to a large body of work studying the phenomenon of online abuse and the communities that it stems from. 
Early work on online harassment focused on cyber-bullying by adolescents in school settings~\cite{quteprints1925}.
Cyber-bullying~\cite{ioannou2018risk,lgbtCyberBullying} has been extensively studied in the context of {\em repeated} hostile behavior towards a well-identified {\em target}~\cite{grigg2010cyber}. Subsequent research has examined cyber-aggression, which generally denotes various forms of harassment (e.g., hate speech, misogyny, racism). %
Our study focuses on understanding cyber-bullying and cyber-aggression in the context of the research world, and how it is difficult to avoid in public-facing careers.

Researchers have also examined cyber-bullying against marginalized communities, as these are disproportionately affected by it~\cite{espinoza2018cyberbullying,zych2015scientific,gardiner2018sa}. 
The canonical incident is the GamerGate controversy~\cite{massanari2017gamergate,mortensen2018anger}, which originated from alleged improprieties in video-game journalism and quickly grew into a larger campaign centered around sexism and social justice~\cite{quinn2017crash}.
Other studies of online abuse against marginalized communities have focused on the Black Lives Matter movement~\cite{ince2017social}, women bloggers~\cite{eckert2018fighting}, and transgender communities~\cite{scheuerman2018safe}. 
We similarly find that researchers who are members of marginalized communities experienced harassment along these vectors.

Our participants also experienced doxing, i.e., the practice of researching and publicly broadcasting private or identifying information about an individual.
Douglas~\cite{doxingConceptual} presents a conceptual analysis of doxing, discussing how it differs from other types of abuse and studying the various modi operandi of doxers. 
Snyder et al.~\cite{IMCdoxing} measures doxing on Pastebin, 4chan, and 8chan -- sites that are frequently used to share doxes online -- studying its prevalence as well as the effectiveness of anti-abuse efforts by social networks. 
We note that doxing is difficult to avoid in the academic setting where office and email addresses are often posted publicly online.

Doxing often leads to coordinated harassment efforts by ad-hoc mobs organized in third-party communities~\cite{bernstein20114chan}.  
Hine et al.~\cite{kekscucks} uncover this phenomenon with respect to 4chan-originated ``raids'' against YouTube videos, while Mariconti et al.~\cite{mariconti2019you} study the characteristics of targeted videos.
Ling et al.~\cite{ling2020first} study the emerging phenomenon of Zoom-bombing, where links to private online meetings are shared online with the goal of harassing the participants.
Hodge and Hallgrimsdottir~\cite{hodge2019networks} discuss the practice of forum raids perpetrated by alt-right communities. 
Marwick~\cite{marwickMMNHPaper} offers a model of organic networked harassment, wherein a single highly connected amplifier incites a mob to pile on.
Many of our participants experienced networked harassment of this form.

For a broader review of online abuse, we refer readers to work by Thomas et al.~\cite{AbuseSOK}, who present a taxonomy systematizing online hate and harassment research. 

\subsection{Remediating Online Harassment}

There have been many studies of remediating online harassment through detection and removal. Automated methods to detect and mitigate online abuse have been proposed on platforms like Twitter~\cite{TwitterAbuse}, Instagram~\cite{Hosseinmardi2015}, and YouTube~\cite{Chen2012DetectingOffensiveLanguage}. 
In particular, various machine learning models have been proposed, with mixed results, to automatically detect hate speech and abusive behavior on social networks~\cite{Chen2012DetectingOffensiveLanguage,nobata2016abusive,Dinakar2011ModelingDetectionTextualCyberbullying,meanbirds}. Manual and hybrid detection and moderation techniques have also been studied and deployed at scale by major platforms~\cite{10.1145/985692.985761,10.1145/3185593,10.1145/3359252}. 

Work intended to expedite the removal of harassment content is complemented by research focused on understanding how socio-technical systems might support those who are the targets of such abuse~\cite{Heartmob} or address the actions of the harassers~\cite{blackwell2018online}. Schoenebeck et al. explore how the concept of justice can be incorporated into the design of such systems, and cultural variations in beliefs about what constitutes a just response to harassment~\cite{schoenebeck2020drawing}. 
A growing concern for researchers' wellbeing is evident in the publication of a guide for avoiding harassment when conducting risky research~\cite{DSguide}; its authors draw on a wealth of research about harassment to educate administrators, supervisors, and researchers, particularly early-career faculty and students, about methods of reducing the likelihood and impact of harassment.
Many of the recommendations aim to help make institutions more hospitable to researchers who may experience harassment, while others are suggestions for improved security and reduced visibility~\cite{DSguide}.
As noted, reducing visibility is difficult early in one's career when professional goals often require visible impact and recognition.

\subsection{Harassment of Academics}

Sexual harassment and discrimination within academia have been found to have a silencing effect~\cite{FITZGERALD1988152}, lead to mental health issues~\cite{doi:10.1300/J013v28n02}, and cause the target of these attacks to leave academia at higher rates~\cite{NAP24994}.
We find that Internet-facilitated harassment can cause the same negative effects. 

Recent work has identified that alt-right groups fixate on and conduct networked-harassment of public researchers which Massanari terms the ``Alt-right gaze''~\cite{doi:10.1177/2056305118768302}.
Some of our participants experienced this phenomenon, which we contextualize in our taxonomy of attacker motivations. Gosse et al. conducted a broader survey of scholars to understand experiences of harassment and the impact it had~\cite{gosse2021hidden}. They report a wealth of descriptive data about the kinds of harassment scholars experienced and the effects it had on scholars, differentiated by a variety of demographic characteristics. Our approach differs from Gosse et al.; however, our findings echo the finding that scholars with marginalized identities are markedly vulnerable to harassment and its harms. 

There are recent first-hand accounts of online harassment which we include in our findings~\cite{firstperson}. Lewis et al. examined the tension of online harassment in journalism~\cite{journalist_harassment}, and Philipp et al.~\cite{Philipp2020} reported on online harassment of social scientists. 
In her recent book, Deo details myriad vectors of harassment throttling female law faculty, with particular emphasis on the career impact of bias from students in a field where teaching is a preeminent consideration for tenure,\cite{deo_2019}
a concern which some of our participants echo.
We include STEM researchers and other knowledge producers as well as social scientists and legal scholars, allowing us to provide a more holistic understanding of online researcher harassment and potentially effective mitigations.

\section{Methods}

To investigate how Internet-facilitated harassment impacts researchers, we conducted 17 semi-structured interviews. 
We chose this method as we were interested in capturing and analyzing the experiences of those who have personally endured harassment as a result of their public contributions to research and knowledge. 
This approach stems from a social phenomenological perspective that acknowledges that the object of social science research is to understand how those being studied interpret and understand their social worlds~\cite{schutz1972phenomenology}.  
By using semi-structured interviews, we were able to collect narrative accounts of what harassment was like for our participants and how they interpreted and responded to the experience. 

\subsection{Participant Recruitment}
Interview participants (n = 17) were recruited using snowball sampling, beginning with personal contacts who we knew had experienced harassment. 
We also reached out to a number of individuals who may not have experienced harassment themselves but could refer us to others who had, given the nature of their research. 
We also looked for news coverage of harassment incidents against academics and reached out to these individuals directly. 
Further, we identified blogs, sites, subreddits, and other fora that listed researchers for the purpose of complaining about their research; some of these actually contain an explicit call to action.%
\footnote{We will not be reporting these for the safety of those targeted.} 
Others provide contact information for the researchers, bordering on doxing, or only names and affiliations (e.g., \url{professorwatchlist.org}). %

Two individuals declined to be interviewed but offered to answer questions via email, while another pointed us to a first-hand account of their harassment experience that they had published on a blog. 
We have reviewed some of these first-hand accounts in (and to inform) our analysis, but do not count them among our participants. Additionally, we have searched for other first-hand accounts individuals have published about harassment experiences, even if we did not contact these individuals about being interviewed. 

\descr{Demographics.} Participants all came from academic spaces and were overwhelmingly directly affiliated with a university. Of these, most were current faculty, though two administrators were also included, as was one undergraduate student. A few participants were activists or journalists, but all of these were actively engaged in research and had close ties to, or backgrounds in, academic spaces.
Some individuals identified themselves in more than one job category. 
As described in Section~\ref{subsec:analysis}, participant recruitment was conducted iteratively as analysis proceeded in order to respond to gaps in representation, or where further data were necessary to fully understand an emerging concept.
Table~\ref{table:profDemo} provides details on the professional status and research areas of participants, while Table~\ref{table:personalDemo} reports demographic information.


\begin{table*}[t]
\centering
\setlength{\tabcolsep}{4pt}
\small
\begin{tabular}{llllllllllllll}
\toprule
\multicolumn{2}{l}{\textbf{Academic Dept}}  &  & \multicolumn{2}{l}{\textbf{Research Area$^*$}} &  & \multicolumn{2}{l}{\textbf{Position$^{*}$}} &  & \multicolumn{2}{l}{\textbf{Country}} &  & \multicolumn{2}{l}{\textbf{Institution$^*$}} \\ \midrule
Computer Science      & 4                   &  & Race/Religion        & 5                   &  & Tenured Faculty           & 6         &  & USA                & 11              &  & Public Univ.                & 9          \\
Social Science        & 3                   &  & Online Communities   & 5                   &  & Untenured Faculty         & 5         &  & UK                 & 4               &  & Private Univ.               & 6          \\
Humanities            & 2                   &  & Crime                & 4                   &  & Administrator             & 2         &  & Other              & 2               &  & Non-Academic                & 3          \\
Law                   & 2                   &  & Gender/Queer Issues  & 4                   &  & Journalist                & 2         &  &                    &                 &  & \textit{\textbf{}}          &            \\
STEM (not CS)         & 1                   &  & Extremism            & 3                   &  & Postdoc                   & 2         &  &                    &                 &  &                             &            \\
N/A & 5 &  &   &  &  & Grad Student              & 1         &  &                    &                 &  &                             &            \\
                      &                     &  & \textit{\textbf{}}   & \textit{\textbf{}}  &  & UG Student                & 1         &  &                    &                 &  &                             &            \\
                      &                     &  &                      &                     &  & Activist                  & 2         &  &                    &                 &  &                             &            \\ \bottomrule
%
\multicolumn{14}{l}{\footnotesize$^*$Some researchers discussed multiple harassment episodes, thus, we include them in all positions, research areas, and institutional}\\[-0.5ex]
\multicolumn{10}{l}{\footnotesize   categories about which the episodes refer to.}
\end{tabular}
\caption{Participants' Professional Demographics.}
\label{table:profDemo}                       
\end{table*}

\begin{table}[t]
\centering
\setlength{\tabcolsep}{4pt}
\small
\begin{tabular}{lrllrllr}
\toprule
\multicolumn{2}{l}{\textbf{Gender \& Sexuality}} & \textbf{} & \multicolumn{2}{l}{\textbf{Race}} & \textbf{} & \multicolumn{2}{l}{\textbf{Ideology$^{*}$}} \\ \midrule
Female                    & 10                   &           & White               & 12          &           & Liberal                 & 8           \\
Male                      & 6                    &           & Black               & 2           &           & Unstated                & 6           \\
LGBTQ+                    & 4                    &           & Other PoC           & 3           &           & Conservative            & 3           \\ \bottomrule
\multicolumn{8}{l}{\footnotesize$^*$Self-identified (in the case of all conservatives) or implied in conver-}\\[-0.5ex]
\multicolumn{8}{l}{\footnotesize sation (in the case of liberals).}\\[-0.25ex]
\end{tabular}
\caption{Participants' Demographic Information. (Note: We did not explicitly ask participants about demographics; this was provided voluntarily as relevant to their harassment experiences.)}
\label{table:personalDemo}

\end{table}


\subsection{Ethical Considerations}
This study was approved by our institution's IRB.
Our participants are already in the public eye, and already experiencing harassment. 
Further publicizing their experiences could bring new waves of harassment, thus, protecting their safety and anonymity was our foremost concern. 
As a result, we do not identify individual participants or their characteristics beyond what is necessary to provide context for a particular quote. 
Some quotes were altered or redacted to mask details. Quotes and attributions have received approval directly from participants, and as a result, have varying degrees of obfuscation per each individual's level of concern for their anonymity.
Demographic data is provided in aggregate. 
Participants were not offered any compensation for participating. 

We are conscious that our work could function as a feedback mechanism for harassers to better understand which tactics are most effective or most damaging. One participant noted this explicitly when reviewing the anonymity practices we would be employing in the study: 
\begin{displayquote}
``So, I don't really want to give the bad guys feedback on whether what they were trying to do worked or not.'' --- \textit{Studying Cyber Crime}
\end{displayquote}
However, as is the case with vulnerability disclosures, we are confident that the discussion of Internet-facilitated harassment and its impact on our participants serves more handily to address the problem at scale than to provide guidance to those engaged in it.

\subsection{Interviews and Analysis}
\label{subsec:analysis}

This paper draws from a series of semi-structured interviews (n = 17) conducted between August and December 2020. Interviews ranged from 40 to 90 minutes, were conducted remotely and audio recorded.
Recordings were transcribed using automated means~\cite{trint}, and the transcripts checked by the interviewer against the original audio for accuracy.  Due to a technical difficulty, one of the interviews was not recorded in its entirety, but the interviewer took detailed notes. 

The first author conducted the interviews and led the analysis of transcripts using an inductive, iterative approach to coding in order to uncover and refine themes in the data, informed by thematic analysis~\cite{braun2012thematic}.  
We wanted to understand harassment of academics, why it was happening, to whom, and what impact it had on academic freedom. 
After the first six interviews were conducted, they were coded by the interviewer, and the emerging themes discussed among the authors. 
Further interviews were conducted and coded using the established codebook and framework. 
After 12 interviews had been conducted, the authors revisited the analysis and made some adjustments to the high-level themes and framework for categorizing experiences. 
At this juncture, the authors also determined that the sample was notably lacking experiences of more junior individuals and those with conservative ideologies, and recruited additional participants. 

After the interviews were complete, the authors had a final round of discussions and made further changes to the proposed framework, to better incorporate the experiences of all participants.

\subsection{Sample Limitations}

Despite concerted efforts to recruit a diverse range of experiences, the sample may be skewed towards individuals with more power, privilege, and seniority. 
Few participants were students at the time of their harassment experience, and more were fully tenured than not.
Knowledge producers at non-academic institutions were all well-established within their fields. 
The lack of untenured individuals might be due to a lack of such individuals engaging in the types of research which may lead to harassment; in fact, participants explained that they did not feel one could do potentially divisive research until they had the structural support (i.e., tenure) needed to do so. E.g.:

\begin{displayquote}
``So the idea that your job is basically unprotected unless you're a tenured full professor like I am... that's another reason why I get to do this stuff. And I didn't until I was tenured in full -- I worked on really boring shit, to be honest.'' --- \textit{Computer Science Professor}
\end{displayquote}
The lack of students in the sample may also reflect a genuine dearth of such individuals engaged in this type of research, or receiving harassment for it. 
Several faculty members we interviewed explicitly indicated hesitance to involve students in work that could incur harassment, while others noted practices they undertake to protect students involved in the work. E.g.:

\begin{displayquote}
``I chose to do [that project] without my research team because I knew it was going to be sticky ... I didn't necessarily want to take any students with me down that path.'' --- \textit{Computer Scientist}
\end{displayquote}
That said, we also spoke with a number of faculty who had themselves been surprised by the harassment they received as a result of their work, so it is perhaps naive to assume that students are not engaged in this work as a result of foresight and sheltering by advisors. It is likely that students are engaging in this work, and being harassed in the process, but not speaking about it openly in a way that would result in a member of our sample identifying them to us, or that students are removing themselves from the work, department, or academia altogether to cope with the harassment~\cite{womenLeaveSTEMHarassment}. 
One participant who was a postdoc at the time of their harassment experience and another who was a PhD student have since become faculty, but indicated that they have intentionally avoided the type of work that brought harassment upon them.
One professor explained observing this phenomenon directly, as their students began to experience harassment:

\begin{displayquote}
``I've had students quit, excellent first rank students, quit the PhD program because they realize they just weren't up to the battle. And, you know, I respect that.'' --- \textit{Queer Art Professor}
\end{displayquote}
Our sample was 70\% white, which is, unfortunately, similar to the population of academia~\cite{nces}. Women are over-represented in our sample compared to faculty as a whole (64\% vs. 31\%), as are members of the LBGTQ+ community (23\% vs. 3\%)~\cite{nces}. 
This is likely because members of marginalized groups are more likely to experience harassment~\cite{sexualHarassmentGradStudents}. 
It is unclear why people of color are not over-represented in our sample. Our non-white participants discussed race as part of their harassment, though none pointed to it as the primary attack vector. 
Women of color spent more time discussing gender issues than race issues. 
As is always the case with snowball sampling, participants are likely to offer referrals to individuals similar to themselves~\cite{snowball}, so it is possible that there are significant groups of experiences we fail to identify. Further research focused on the intersection of race and harassment in the academic context could serve to illuminate this issue.

\subsection{Positionality}
Authors invariably bring a set of assumptions, beliefs, and experiences to their research. The authors of this paper vary in their methodological expertise and orientation; however, in this study, we adopted an interpretive, phenomenological approach to understand the experiences of people who have experienced harassment as a result of their contributions to public knowledge. In order to gather wide-ranging experiences that could yield a robust characterization of harassment and its effects, we actively sought out a variety of political perspectives. We attempted throughout the data collection and analysis process to remain cognizant of our own political beliefs and how they may influence our analysis. Although each author holds individual beliefs about the moral values associated with specific harassment incidents, this paper does not explore the concept of ``justified harassment,'' nor attempt to adjudicate in what contexts harms might be ``deserved.'' We direct readers to scholarly literature on when online harassment is perceived as justified for an excellent discussion~\cite{blackwell2018online}. Finally, all authors have personally experienced varying degrees of harassment as a result of our work; while these experiences have undoubtedly influenced our interpretations of data, the analysis itself is limited to the interview data collected.%

\section{Harassment Motivations and\\ Enactments}

Participants' harassment experiences ranged broadly in terms of their harassers' (perceived) modi operandi, ideology, and content. Our participants could not always identify the individuals and groups harassing them but generally felt that they understood why they were being harassed. Through iterative analysis, we began to link the suspected motivations for harassment across participants' stories with features of enactment. We explain the main categories of suspected harasser motivation we observed, the type of harassment each manifests, and the research (and researchers) targeted by each. 

Our participants were victims of harassment, not the harassers themselves; therefore, our access to harassers' motivations is based on perceptions and indirect communication. In many cases, our participants' harassers communicated motivations as part of their harassment; in all cases, imputed motivations were based on actual communications from harassers and our participants' knowledge of the context in which the harassment took place. 
Indeed, most descriptions of harassment experiences included some statement of motivation or justification on the part of harassers, often, these statements were the totality of the harassment experience, e.g. tweeting that ``You're a terrible person for doing this research'' is both an act of harassment and an explanation of motive.
Some participants postulated that the \textit{true} motivations of their harassers were different from the stated motivations, often seeing the stated motivations as dog whistles and the true motivations as bigotry.
This is a well-documented phenomenon in political expression, \cite{Searles_2016}
and for participants studying issues of race and gender, it is reasonable that as experts in the field, they recognize veiled racist, sexist, or homophobic statements.
Therefore, in addressing these types of experiences we have differentiated between harassment which is \textit{patently bigoted}, i.e. the harassers are openly stating bigotries, and \textit{`political objections'}, wherein we discuss participants' perceptions of the genuineness of harassers' stated objections.

In addition to participants' direct statements, we used the following analytic criteria to distinguish between different motivations for harassment: the \textit{visibility} of the harassment, %
how \textit{targeted} it is, %
and how \textit{personal} in terms of its focus on identity-related characteristics.
An extension from visibility is whether the harassment is \textit{networked} -- often, harassment that happens in public spheres follows a model outlined by Marwick \cite{marwickMMNHPaper}: a public figure expresses a grievance with someone, and a pile-on ensues. 
Whether or not harassment is networked is, like our analytic criteria, a property of the harassment, not the motivation. 
However, it is also a deliberate tactic and presents a set of threats and constraints orthogonal to its underlying motivation, thus, we discuss the categories of motivations, and then networked harassment as a whole contextualizing it to each motivation in order to minimize repetition.

We posit that there are three primary motivations for the harassment of public figures like researchers: 
1) \textit{Self-Preservation}, in that the research poses a real or perceived threat to the harasser 
2) \textit{Ideology}, when the research is offensive to the harasser or challenges their ideology; 
and 3) \textit{Performative} harassment for personal or social gratification.
Although we believe these categories are useful in understanding specific instances of harassment, we note that many of our participants experienced harassment in more than one of these three categories. 
One type of harassment may also lead to another; e.g., researchers experiencing ideological harassment from powerful entities may then experience self-preservation harassment from their institution. %

\begin{figure}
\includegraphics[width=0.49\textwidth]{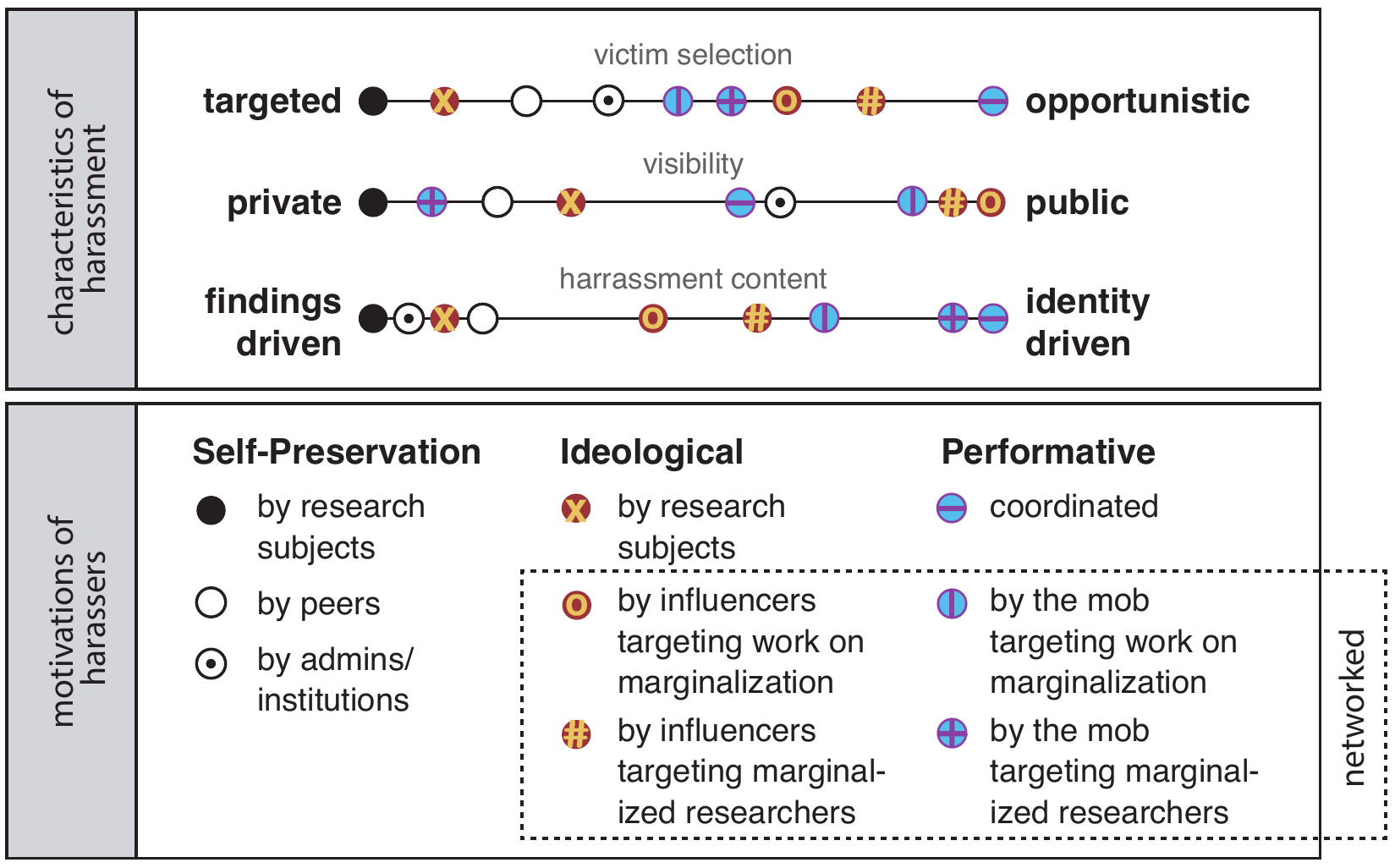}
\caption{Axes for Distinguishing Three Harasser Motivations: Self-Preservation, Ideological, Performative
}
\label{figure2}
\end{figure}

Figure~\ref{figure2} provides a visual representation of the analytic criteria described above and their relationship with the forms of harasser motivation described in the remainder of this section. For example, the figure depicts axes like visibility---certain types of harassment manifest privately, with the harasser communicating one-on-one with the researcher. Other forms manifest publicly, with harassers condemning the researcher on social media. Some are in spaces in-between: pseudonymous fora such as 4chan, blog comment sections, closed-door meetings, or phone calls to one's department. In the figure, one can see trends, for example, that harassment stemming from self-preservation tends to be more targeted, private, and driven by findings or content whereas performative harassment tends to be more opportunistic, public and driven by identity characteristics of the targeted researcher. Note that this visual representation necessarily simplifies aspects of the harassment our participants experienced. In some cases, we noted where most harassment activities were located but acknowledge that idiosyncratic experiences may drift from these loci. For example, while influencer-incited mobs usually manifest as networked harassment publicly on social media, we did observe examples---generally in the case of politically progressive research---of an internet-instigated crowd sending private letters to individuals and calling their departments.

Additionally, we provide Table~\ref{table:huge} as an Appendix with a complete enumeration of our participants' harassment experiences.

\subsection{Self-Preservation Harassment}
\label{subsec:self-pres}

Self-preservation harassment of researchers is driven by fear; an entity or individual perceives that research poses a threat to them. 
For instance, when studying criminal or clandestine %
groups, harassment often comes from the subjects of the research, wherein the very idea of being observed threatens their secret status. 
Researchers advancing potentially controversial ideas may also face this type of harassment from administrators who perceive it as a threat to funding sources, and colleagues who see it as a threat to the legitimacy of the field. 
The primary goal of self-preservation harassment is to stop the research.
As such, it is the most persistent of the three forms, and the least identity-driven---it doesn't matter who is doing the research, it must be stopped. 
It poses serious financial, logistical, technical, and physical risks, but participants often reported lesser emotional consequences. 
This harassment is generally private, targeted, and impersonal.

\subsubsection{Harassment by Subjects of One's Research} 
In the last decade, illicit groups have moved portions of their operations online, making them more vulnerable to observation~\cite{fakeAntivirus, spamValue, priceless, muleScams}. Researchers studying these groups may be harassed by the group itself or individual members.

Criminal enterprises and terrorist organizations posed strong threats, leveraging the resources of the organization to engage in harassment. 
One participant described how an organization they were studying had used its own bad reputation to incite harassment of those studying them:

\begin{displayquote}
``A release came out by an extremist organization in %
the Middle East talking about all of these new %
people who had joined the group and they named them. [The people] were actually all researchers of violent extremists, but just like one vowel in the name had been changed.'' --- \textit{Studying Extremism}
\end{displayquote}
These organizations perceive that the research may aid law enforcement in locating them, cause financial institutions to stop engaging with them, or otherwise interfere with their operations. 
This type of harassment is clearly connected to the research and is often directly attributed to the group being studied. In some cases, threats are signed and provide a clear directive to stop the research. 
When asked how they knew who was harassing them, one participant explained:

\begin{displayquote}
``You're going to get a direct message on Twitter that says, `You or something you care about, it's going to suffer if you don't stop.''' --- \textit{Studying Cyber Crime}
\end{displayquote}
While some organizations are not threatened by research, their members may perceive a personal threat, for example, if they fear their affiliation with the group being exposed. 
For example, David Duke\footnote{David Duke is an American neo-Nazi, convicted felon, and former grand wizard of the Knights of the Ku Klux Klan \cite{david_duke}} 
may not be bothered by research on the Ku Klux Klan, while others may fear that being outed as Klansmen will ruin their reputation or cause them to lose their job. 
A professor studying alt-right groups indicated that harassers are motivated by fear for their livelihoods:

\begin{displayquote}
``I'm not going to sit there and be on the phone all day, you know, it's ridiculous. But that's what they imagine I'm doing, is like, calling the pool company and getting pool boys fired.'' --- \textit{Studying Alt-Right}
\end{displayquote}
Self-preservation motivated harassers are the most pernicious, continuing to escalate if their initial threats do not serve to stop the work in question. 
This harassment may lead to physical violence, weaponize law enforcement through SWATing\footnote{Calling the police to someone's address on account of something serious like a bomb threat, aiming for SWAT deployment.}
 attempts, or be technologically sophisticated, including spear-phishing campaigns\footnote{A highly targeted version of phishing, aiming to trick someone into disclosing their credentials.}, doxing, %
or DDoS attacks.\footnote{Distributed Denial of Service: attempts to inhibit function of a Web service by overwhelming its servers with traffic.}
Individuals experiencing this type of harassment have legitimate concerns for their physical safety at work and at home; three participants report receiving outreach from the FBI warning them of a threat to their life.

Overall, five participants experienced self-preservation harassment from the subjects of their research.
While this harassment is highly targeted, it is less identity-driven. Women experiencing self-preservation harassment reported fewer gendered attacks than women experiencing other forms of harassment. 
 A professor studying conspiracy theorists was surprised by this:

\begin{displayquote}
``I actually don't know why, but I haven't received a lot of misogynistic content, and I've been surprised that there hasn't been. Most of it has been about the work I'm trying to do and trying to undermine that, as opposed to trying to do something more personal.'' --- \textit{Studying Conspiracy Theorists}
\end{displayquote}
Gender may be an easy attack vector for trolls, but does not appear to be as readily weaponized by self-preserving attackers who are motivated, sometimes criminals, and willing to escalate to more serious methods of attack.
These harassers often directly stated their motives, and when they did not, the fact that they were the subjects of research which did not present them favorably supports participants' ascription that it was motivated by self-preservation.

\subsubsection{Harassment by Colleagues and Administrators} Researchers who challenge their own field may find that peers perceive their work as an affront to the discipline---e.g., when women in STEM disciplines write about gender bias or sexual harassment in their own fields. Certain corners of the Computer Science community have perceived the study of algorithmic bias as a fundamental threat to the credibility of their work, disputing that such work should be undertaken, or ostracizing researchers studying these issues, as seen recently with Dr. Gebru at Google~\cite{Gebru}.  
A professor studying demographic biases in language models received hate mail from colleagues to that effect:

\begin{displayquote}
``[One] note that I received was against the idea of fairness and measuring bias for fairness. Like, they were against the problem itself, which is hard to understand for me.'' --- \textit{Computer Science Professor}
\end{displayquote}
The interplay between this self-defense reaction and the latent misogyny and racism present in places like 4chan~\cite{kekscucks} can mean that sects of the Computer Science community go beyond dismissing peers' work and engage in outright harassment, as discussed later in Section~\ref{subsec:marginalizedprofs}.

Participants in Computer Science and adjacent fields said they were concerned about receiving backlash when undertaking projects that might present findings that were at odds with the popular views in the field, for example, with respect to entities like WikiLeaks or Anonymous. E.g.:

\begin{displayquote}
``I was aware that we might become critical of [things] that other academics had historically been very positive about, for instance, WikiLeaks. I had this worry: `I'm going to be talking about something that's controversial, and many people in academia may not agree with me.''' --- \textit{Computer Scientist}
\end{displayquote}
In some of these cases, harassers' motivations were stated explicitly: they felt the research was unethical or invalid. Participants postulated that their reason for seeing it as such was because it challenged the legitimacy of their work or an accepted orthodoxy. Participants noted that these objections, expressed in emails, blog posts or letters could be quite lengthy, and include a handful of valid critiques of their methodology amongst a large amount of personal, often bigoted, attacks.

Harassment can stem from an administration as well as peers. Researchers whose work pushes boundaries may find that administrators perceive their work as a threat to the future funding of their institution or department. 
While academic administrators as a group are overwhelmingly left-leaning~\cite{abrams_2018}, faculty advancing social justice causes at institutions with conservative donors may face professional censure~\cite{flaherty_2020} or censorship of their work. One participant's work was censored after a Republican politician threatened the institution's funding:

\begin{displayquote}
``They broadcast all sorts of threats on Fox News, including that they would remove 10\% of [the museum's] budget if the exhibition wasn't closed down. The secretary of [the museum], who was a man with a habit of knuckling under to conservative fire breathers, said that he saved the exhibition by censoring the one work that they found objectionable.'' --- \textit{Queer Art Professor}
\end{displayquote}
By the same token, administrators may perceive outspoken conservative scholars to pose risks to the reputation or legitimacy of their institution.
This may result in direct harassment, such as firings or workplace hostilities~\cite{cadf}, or indirect harassment through failing to support these individuals when they are harassed by other parties or failing to condemn harassment on part of the student body~\cite{abrams_2019}. 
The leader of a conservative student organization was disheartened that her university had failed to respond to a Zoom-bombing attack in the same way it had to a similar attack on the school's Black Student Alliance earlier in the pandemic:

\begin{displayquote}
``I was upset because this [other] student org had suffered an attack like ours, and the university had issued a statement and an email to the entire school condemning the attacks. I was expecting the same sort of thing, and that never happened. Theirs happened first, but more importantly, it was a dialog on racial justice, and obviously the political climate we're living in---that was a horrible thing to happen and I feel so bad for them. But then, when I look at our event, it was very similar, because it's a women's group, and we were trying to talk about sexual violence." --- \textit{Conservative Undergraduate}
\end{displayquote}
In cases where administration and donors are directly responsible for harassment, the motivation is made explicit in the ultimatum presented; you shut down the objectionable project or you lose funding. 
In other cases, victims of harassment are simply left with the impression that their institution \textit{chose} to do nothing, silently siding with the harassers. 
While this cannot be proven, responses from administration after high-profile incidences of harassment by students against conservative faculty lend credence to this perception. 
For example, in the case of Prof. Samuel Abrams, the administration first responded to student demands and condemned violence and vandalism on their part only after being pressured. \cite{abrams_2019}
The authors reiterate that we do not seek to pass judgment on what harassment is justified or merited; Dr. Abrams perceived his politics resulted in his school's unenthusiastic defense, and this seems to be a reasonable interpretation.

Three participants experienced discriminatory treatment or direct harassment from administrators or department chairs. 
A further seven expressed dismay with the way their organizations handled harassment against them. 

\subsection{Ideological Harassment}

This type of harassment stems from strong, genuinely held beliefs that certain research is immoral or illegitimate, %
or that certain individuals are incapable of doing research, or should not hold positions such as professorships.
Ideological harassers may be well-defined groups, like hate groups, or motivated individuals. 
The primary goal of this type of harassment is to discredit or undermine the research or the researcher, thus vindicating or reinforcing the harasser's ideology.
Even if in bad faith, this harassment engages with the \textit{nature} of the research. Gamergate is a well-known and studied example of ideological harassment~\cite{massanari2017gamergate,salter2018geek}.

Ideological harassment creates professional difficulties for researchers, either through direct interference with their work or employment or due to overwhelming logistical and emotional labor required to remediate the harassment. 
Harassers of this form often attempt to cause reputational damage to researchers, seeking formal censure, generally through the loss of a position, denial of tenure, or similar, or to create comparable career impacts by generating public outrage, creating a disincentive to employ, associate, or collaborate with them, often referred to as `canceling' someone.

The authors acknowledge that `cancel' is a loaded phrase, and do not seek to weigh in on which `cancelations' are justified --  certainly \#MeToo \footnote{\#MeToo is a movement against sexual abuse and harassment involving publicizing allegations of sex crimes and harassment.} 
is evidence that active public campaigns can be necessary to remove predators from an industry. Rather, we note that `cancel' is an increasingly common term of art representing a particular kind of censure and consequence, and that calls to harass or delegitimize someone are often made explicitly in these terms. 
Concerns about being `canceled' were echoed by our participants across political and ideological lines.

Ideological harassment is targeted and may occur in public or private; it is not identity-centered, though identity may be weaponized as part of the argument for discrediting the work or when bigotry is itself the underlying ideology. 
Organizations like media outlets, generally those with strong political leanings, may engage in and encourage this harassment by writing inflammatory articles, and ideological influencers may use their platforms to trigger networked harassment. 
We discuss this further in Section~\ref{subsec:networked}.

\subsubsection{Objection to Being Studied} 
Unlike in self-preservation harassment, some groups under study do not perceive the research as a threat and do not necessarily want the work to stop, they simply do not {\em want} to be studied.
For example, %
conspiratorial ideologists may object to research about them as being inaccurate or immoral--- another example of the deep-state determined to hide the truth---but perceive the publicity the research brings as an opportunity instead of a threat. 
One researcher highlighted this duality when studying Anonymous prior to their pivot to hacktivism:

\begin{displayquote}
``Anonymous -- in the trolling, pre-hacktivist era -- had a little bit of a fight club ethic\footnote{A reference to the 1999 movie Fight Club -- ``the first rule of fight club is you don't talk about fight club.''}, they chided outsiders or journalists but their trolling was also executed to land attention. So journalists were baited to cover their trolling but if you tried to examine them in other terms, you were accused of ruining the Internet and /b/\footnote{Anonymous grew out of the /b/ board on 4chan.} by trying to understand what was going on in this world.'' --- \textit{Studying Anonymous}
\end{displayquote}

The harassment these groups engage in is mostly trolling: online, over the phone, and through the mail. 
Much of this trolling is private; participants described receiving slurries of forwarded junk mail or being signed up for hundreds of listservs. %
Harassers may also post hateful comments on Twitter, on YouTube videos of one's talks, or in their own fora, where they know the researcher will end up reading them in the course of observation. 
Several researchers studying adversarial online communities noted that, by studying an information space, they inherently became a part of it. 
One individual, studying 4chan, noted that much of their harassment was written on the boards they studied, %
making it easy to avoid the harassment but difficult to do their work:

\begin{displayquote}
``The threats that I saw were mostly on 4chan. They write, `hey, we know you guys are studying us. We know you are looking at our posts. We have written this for you.' And then they write a hateful thing, or the threat. ... I study 4chan, but 4chan is not a place I like to see. I don't hang around there because, well, it's a toxic place, generally speaking.'' --- \textit{Studying 4chan}
\end{displayquote}

In these cases, harassers' motivations are explicitly stated as a preface to harassment which is somewhat agnostic to the researcher as an individual: you are studying 4chan, we (4chan) do not like being studied, therefore we will drop-ship you a box of plastic frogs.
Four participants experienced harassment from the subjects of their research which was not motivated by self-preservation. 

\subsubsection{Objection to Work on Race, Gender, and Marginalization.} Research that addresses issues of race, gender, and other aspects of marginalization is likely to incur harassment by individuals and groups whose ideologies are offended or challenged by the work. 
This takes two main forms: 1) private, targeted harassment from those who openly hold bigoted ideas and are offended by the work, and 
2) public, opportunistic harassment from those who feel the work accuses them of having bigoted ideas. %
A geneticist studying the disconnect between scientific and popular understandings of race explained the divide between these two types of harassers:

\begin{displayquote}
``The concept of race being a social construct is completely non-controversial among geneticists and anthropologists and scientists more broadly. [...] There are both people who are active, self-identifying racists and people who don't consider themselves to be racist, but are expressing racial stereotypes. And to a certain extent, my work confronts their views.'' --- \textit{Geneticist}
\end{displayquote}

\descr{Patent Bigotry.} Individuals and groups that hold -- and acknowledge -- beliefs we might classify as racist, misogynist, or homophobic  %
reached out to our participants directly to inform them that their findings or conclusions were incorrect.
These harassers write anonymously, not willing to be tied to their beliefs. Five participants experienced harassment of this form.
Openly bigoted harassers may not dispute the accuracy of research demonstrating or measuring marginalization, instead asserting that that is the correct or natural order of things. 
Harassers responding to work on gender bias in algorithms did not dispute the findings, but rather felt they were not findings at all:

\begin{displayquote}
``Many people rejected the findings. They were like, `of course women will be associated with family and arts and men with career and science' [...] or saying that women are weaker and inferior, and of course it should be like this, and I'm just trying to change this idea.'' --- \textit{Computer Scientist}
\end{displayquote}

Research that seeks to undermine bigoted ideas, such as the geneticist demonstrating that race is a construct, may instead face objections that the work is invalid, merely trying to advance the social standing of a marginalized group, a prospect to which the harassers object intrinsically. 
In the case where the researcher is themself a member of the marginalized group, harassers leverage this as `evidence' that the research is invalid: 

\begin{displayquote}
``I'm mixed race myself. And that's something that comes up, and it's sometimes used as a weapon to say `well he would think that because he's [mixed race].' And so, here's another race warrior who is bastardizing science because of his personal background.'' --- \textit{Geneticist}
\end{displayquote}

This harassment may seek to interfere with the `objectionable' research directly; for example, a survey of trans students received responses in which individuals claimed to sexually identify as an apache attack helicopter, undermining the usefulness of the results.
In these cases, the motivations of the harassers are either explicitly stated or clearly implied by the presence of slurs and vitriol in the content of the harassment; it's about race, gender, etc. and they're not pretending it isn't.

\descr{`Political' Objections.} %
Individuals who do not believe themselves to be bigoted, but find that research challenges their views, or feel it implies they are bigoted, 
often object loudly and publicly to this work. 
Participants believed that this harassment stemmed from underlying bigotry, but harassers with political objections may not perceive this to be true of themselves. 
They tend to use the language of political bias or political correctness, assert (often scientifically invalid) counter-examples or counter-narratives, or dismiss the work as invalid for some other reason. 
This generally happens in public spaces and is opportunistic, beginning after the work has received media attention; this may be media that is supportive or antagonistic to the work, but we noted that harassment typically starts after a high-profile individual circulates the story, leading to networked harassment. 

Research focused on issues of marginalized people within the humanities and social sciences, for example, feminist interpretations of classic literature may be outright dismissed as not `real science.'
Departments focused on the experiences of marginalized people, such as Queer Studies, may face objections to their legitimacy as fields of research, though harassers tend to target specific researchers, generally those most prominent.
Two participants discussed this issue.

Other participants faced harassment that challenged their work by attacking the validity of the premise, the legitimacy of the findings, or the impartiality of the researcher. Harassers who choose to engage with the research to some degree often make broad complaints about the ideology motivating the research, challenging its premise, or the legitimacy of academia as a whole, often implying the work is biased or entirely fabricated. These harassers may use information about the researcher's institutional affiliations, sources of funding, or information from their social media profiles as `evidence' of bias. These complaints may come via private channels, such as letters, but are often lodged in public online spaces. Nine participants discussed seeing accusations that their work was part of some grand conspiracy and therefore not to be trusted. Objectors from the right declaimed research as part of the `deep state' or `liberal agenda,' while objectors from the left discussed colonialism, the CIA , and the military-industrial complex. E.g.:

\begin{displayquote}
``They took a screenshot of my funding - I have grants [from the defense space], so I must be a US government shill or something. I was like, you just screenshotted my webpage.'' --- \textit{Computer Scientist}
\end{displayquote}

\begin{displayquote}
``They'd say: `this research comes from liberals, it represents the ideologies of liberals, universities shouldn't be like this, they shouldn't be doing this research' and so on.'' --- \textit{Computer Science Professor}
\end{displayquote}
Some engaged with the work more directly, challenging the findings via counterpoints -- often widely held beliefs and misconceptions -- or arguments that fail to understand the nuance or ambiguities involved in the research. Four participants faced this type of objection. Not all those who reach out in this capacity are harassers; many people reach out in good faith, hoping to reconcile their understanding of the world with research that confronts it.
Participants were willing to engage with this outreach, but also noted that it can be polite on its surface, but sufficiently combative or obtuse that they begin to question whether it is genuine. A professor whose work confronted racist narratives based on poor-quality evidence explained:

\begin{displayquote}
``I've explained so many times that it's just not possible to give an accurate figure and there's no data collected that would allow you to. We just don't collect it that way. So people go `you've said [this] isn't the right figure, but you're not even giving us a figure.' And well, that's the point. There isn't one. I don't know whether they've read it and just don't really understand the basics of science or math or literacy, or whether they haven't read it and are just trying to push this line.'' --- \textit{Tenured Professor}
\end{displayquote}
Varied harassers fall into this category: individuals who are not directly affected by the work but are offended by it for a reason they feel is political. This includes genuine political grievances, dog whistles, conspiracy theories, and others.
As such, participants' ascription of motivations were varied and nuanced for these types of harassers. 
This harassment, like that in most of the other categories, often stated in its content what ax the harasser had to grind, however, participants had various opinions on whether these stated motivations were true and genuine, or whether there was some \textit{deeper} motivation. 
Psychological, sociological, or anthropological research could follow up on this,  seeking to establish the \textit{true internal motivations} of these harassers at more intimate levels. 

\subsubsection{Objection to Marginalized People Being Researchers} 
\label{subsec:marginalizedprofs}
Members of marginalized groups who are researchers may face harassment on account of their identity, as a basis for questioning their credibility or competence, or from those who are simply bothered by the idea that \textit{someone like them} is a well-respected researcher. 
This harassment is ideological, but orthogonal to one's research; the ideological objection is to their existence in their field, or in academia more broadly, or simply to the fact that they have any degree of power, influence, or financial success.   

Marginalized scholars may face this type of harassment from within their own communities, particularly when those communities have non-academic practitioners, e.g., Law or Computer Science. 
Female participants in both fields reported receiving deeply misogynist harassment objecting to their existence or prominence, which they identified as coming from members of their field but from outside the academy. 
A Law professor recounted harassment from legal practitioners:  

\begin{displayquote}
There was an economic downturn in 2008, so a lot of people that graduated from law school in 2009-11 had a lot of debt and couldn't get jobs. So there was this large group of angry law school graduates who felt like they'd been scammed. They looked at me, and I think they saw somebody who was perpetuating the scam. Maybe the way to understand my situation is that there was this built in audience that was already prone to engage in this type of behavior and because of their specific situation, found me an appealing target. --- \textit{Law Professor}
\end{displayquote}
Marginalized scholars in all fields may face harassment simply on account of their existence, as do other visibly successful people~\cite{famous_harassment}. Participants attribute this harassment to a genuine belief that certain groups are `unfit' to be faculty or a sense of entitlement from traditionally privileged classes. This harassment uses the language of propriety or moral standing, or discusses affirmative action, and happens in public spaces, such as Twitter replies or YouTube comments. A Black professor explained that her harassers simply hate that she has success:

\begin{displayquote}
``I am very openly Black and pro-Black, queer and a woman and disabled. I'm also very sex-positive and sex worker positive. And it's all these things that are terrifying to these folks. It's all things that they hate, right? And they're like, `how is this person, who operates within all these identities that we have been told are bad and make someone less valuable, in this position of power and making all this money? Certainly, something has to be wrong here.' It's not my research that's being challenged. It's just like literally how I exist in the fucking world.'' --- \textit{Tenured Professor}
\end{displayquote}
Because of the latent bigotry in this type of harassment, participants were incredibly clear about the motivation behind it, even when it was not stated as explicitly as `you don't belong here,' discussions of affirmative action made it clear to participants that the harassers felt that by virtue of their identity, their existence in the space they occupied was illegitimate. 

\subsection{Performative Harassment}
Self-preservation harassers aim to \textit{stop the research}, fearing its impact upon them, while ideological harassers aim to \textit{discredit the research}, seeking to validate their own ideologies. 
For performative harassers, harassment is both the end and the mean. 
It is untargeted and opportunistic; the harassers are not concerned with the details or even the nature of the research and do not seek out targets. 
Instead, they respond to others' complaints by piling on. 

The purpose of this harassment has little to do with its target; it is either to make the harasser feel good by putting others down, as is understood to be a primary motivation of bullying~\cite{smith1999nature}, or it is the hope that someone whose opinion they value, either a specific individual or a broader community, will see them engaging in the harassment and praise them for it.
This harassment often does not engage with the research itself and tends to be personal with respect to identity traits of the researcher, as these make for easy and effective attacks. 

This mostly happens in public, often manifesting as networked harassment. %
While much of this harassment can be described as trolling, it is critical to understand that it can escalate to other forms of harassment and real-world violence, and thus, should not be dismissed or brushed off.

\subsubsection{Coordinated Performative Harassment}
Certain online communities, such as 4chan, engage in consistent acts of harassment in a manner that is public and often identity-focused, but opportunistic. 
Harassment is organized informally on those boards, 
with users encouraging each other to engage in harassment, and commending each other for having done so~\cite{kekscucks,mariconti2019you}.
Zoom-bombing, which happens relatively privately and can appear to its victims to be targeted, also emerges from these boards opportunistically, for the `fun' of it~\cite{ling2020first}.
The motivation of this behavior is likely self-gratification through bullying and seeking praise from peers and acceptance from a community. 

While some communities claim specific ideologies (e.g., racism, misogyny, antisemitism), these beliefs may or may not be genuinely held. The ideologies in these spaces are so toxic that they may be proffered simply for their toxicity, as it makes good ammunition. 

This said, there are surely members of these communities who hold these beliefs genuinely, and who become radicalized in these spaces, leading to serious, violent outcomes. 
For example, shooters in Christchurch and El Paso, killing over 70 people combined, posted live updates on 8chan, eventually leading to the platform being taken down~\cite{8chan}.
4chan, of course, is also the birthplace of Anonymous, a group with significant technological capabilities. 
Harassment communities can trigger networked harassment, but often large-scale attacks are still perpetrated only by members of the community, without ginning up outside support.
Thus, these are coordinated, but not networked.

\subsubsection{Organic Performative Harassment}
\label{subsec:operf}
This is influencer-driven harassment: a particularly high-profile individual calls someone out, leading their followers to commit harassment \textit{en masse}.
It is organic because the followers did not all collect in a private space and \textit{decide} to harass someone, they simply piled on overtime. 
The influencers often do not commit harassment directly, instead relying on their devotees to do it for them and avoiding Twitter bans or other ramifications, even if they know full well that they are inciting a pile-on. 
Motivations of the influencers aside, the motivation of the mob -- parroting false or reductive claims about the research, or simply throwing slurs at the researcher -- can be understood as seeking to gain praise from the influencer and others who follow them and is thus performative. E.g.:

\begin{displayquote}
``But a lot of times it's people who are on the periphery of these distinct groups [...] it's not necessarily somebody who I've directly exposed or written about. And very often it's just somebody in the community that maybe wants to appear to be a badass or they're trying to curry favor with somebody who's a perceived influencer in that group.'' --- \textit{Journalist}
\end{displayquote}
This type of harassment is almost exclusively networked, so is discussed further in Section ~\ref{subsec:networked}.

\subsection{Networked Harassment}
\label{subsec:networked}
We observed several distinct types of networked harassment, 
all of which could be described by Marwick's theory of Morally Motivated Network Harassment (MMNH)~\cite{marwickMMNHPrez,marwickMMNHPaper}, in which an amplifying node alleges an individual has committed a moral transgression, and then their network harasses that person. In defining MMNH, Marwick discusses seeing this phenomenon in political and apolitical areas, including schools of thought on interior design~\cite{marwickMMNHPaper}. 
In our sample, the influencers tended to be journalists, politicians, or thought leaders in political spaces. %

We identify some additional layers of motivation to which her framework is agnostic: 1) the motivation of the \textit{amplifier}, as interpreted based upon the \textit{truth} of the allegation, and the \textit{genuineness} of the outrage, and 2) whether the attacks are coordinated.
The intersection of these factors results in harassment which is markedly different in its form, content, and objective.
We observe performative harassment which is networked, ideological harassment which is networked, and self-preservation motivated attackers who use ideologically motivated networked harassment as a tactic.

\subsubsection{Coordinated Networked Harassment.}
Our participants experienced straightforward examples of networked harassment emanating from ideological communities after a highly networked member of the community writes something about a researcher or their work in a dedicated forum for the community. %
The claims made about the researcher are generally framed in a defamatory manner, but true, and may pertain to the researcher's work or some aspect of their identity. 
For example, a white supremacist group harassing a Black researcher has made an accurate claim (that the person is Black) and has a genuine belief (racism), thus, we understand the amplifier's motivation to be ideological. 
These writings often encourage the community to harass that individual, either explicitly or implicitly. 
The harassment which manifests from these networks is often not visible to the broader world, such as hate mail or phone calls. Because this harassment is private, we interpret the motivation of the network to be ideological as well. E.g.:

\begin{displayquote}
``They would say things like `I'm writing in response to your anti-American, kike comments [I saw in] the Catholic League newspaper.''' --- \textit{Jewish Professor}
\end{displayquote}

\subsubsection{Organic Networked Harassment.}
Often, harassers seek to weaponize broader societal and cultural flashpoints. 
The accusation, amplification, and harassment happen in public, mostly on Twitter: an influencer or media outlet writes something about the research or researcher that is defamatory, often less than true, and highly inflammatory to a political cohort. 
They are thus able to mobilize a broad and loose coalition of harassers. Amplifiers have varying motivations for inciting the mob; in all cases, we interpret the motivation of the mob to be performative.

\descr{Ideological Motivation}
Amplifiers with a \textit{genuine} objection to the work or researcher may seek to incite a mob to reinforce their ideology. These allegations may be true but are generally nebulous, reductive, misleading, or patently false. 
For example, calling someone an ``SJW'' (Social Justice Warrior) opens them up to harassment from a large group of people with fairly disparate ideologies, and little sense of what the researcher actually studies. Nine participants had this experience; all had amplifiers that were right-leaning, two also had amplifiers that were left leaning.

\begin{displayquote}
``[The amplifier] is a Twitter follower of mine or perhaps a non-follower to whose attention my tweet has been drawn. So he furiously tweets out, `this guy's a Nazi' or `this guy's a communist' some absurd allegation or characterization. Wherever that individual happens to be coming from -- right or left -- he denounces you as the opposite.'' --- \textit{Conservative Legal Scholar}
\end{displayquote}

\descr{Self-Preservation Motivation.}
Noting the cultural relevance and career implications of someone being `canceled,' harassers motivated by self-preservation may seek to stop inquiry into their affairs by manufacturing scandal, and therefore ideological networked harassment.
These amplifiers create completely false narratives about researchers or take things wildly out of context.
This may involve the harasser weaponizing their own bad reputation, and portraying the researcher as one of them -- for example, someone studying, and therefore engaging with, the alt-right may be falsely outed as a sympathizer, in the hopes they are ``canceled'' by the left. Participants who were experiencing self-preservation harassment indicated that this threat was a primary concern:

\begin{displayquote}
``Anthropologists do often take, not necessarily a sympathetic stance, but they are trying to understand what's going on on their own terms. I could imagine something where they try to portray me as being, like, sympathetic to this world, and that would be kind of a nightmare.'' --- \textit{Anthropologist}
\end{displayquote}

\descr{Financial Motivation.}
Lastly, in some instances of networked harassment, the amplifier has no interest in the research, but rather chooses a target opportunistically, seeking to gain from the pile-on financially or politically. Accusations may or may not be true, and the amplifier's objection is disingenuous. Amplifiers are often fringe media outlets seeking to drive traffic to their websites---which translates directly into ad revenue---by drumming up outrage. Participants also discussed politicians seeking to drive voter turnout, and influencers looking to bolster podcast listenership. Six participants discussed amplifiers of this form.

\begin{displayquote}
``A right-wing click farm will find my article and come up with [an inflammatory headline]. Then they put that on Facebook and it goes viral on there and they get like tons of clicks. One guy just sells supplements. He's like a Great Value Alex Jones\footnote{Alex Jones is an American far-right radio host and conspiracy theorist. \cite{alex_jones} He sells supplements.}
 or something. --- \textit{Studying Alt-Right}
\end{displayquote}
It is difficult to say for certain that those inciting these virtual mobs are motivated by financial or electoral gains, but participants -- sometimes experts in fringe information spaces -- made informed suggestions. Future work could investigate financially and electorally motivated networked harassment, inside and outside of academia, perhaps in the capacity of a financial network.

\section{Unique Impacts and Constraints in Academia}
In this section, we discuss the personal, emotional, and financial consequences our participants experienced as a result of their harassment, as well as the professional ramifications which were a direct result of the harassment. 
We also focus on the ways in which harassment makes it difficult to continue one's research and ways in which academia and similar fields make individuals uniquely vulnerable to harassment and impose constraints on remediating it. 
Finally, we discuss the impact of harassment as a phenomenon on academic freedom.

\subsection{Impact of Harassment}
While the individual impacts of harassment are severe, they are also well-studied~\cite{quteprints1925}. 
Targets of harassment experience emotional distress which can be severe, and may experience social isolation, either as a direct result of harassment or as a consequence of removing themselves from spheres in which they are harassed. In more serious instances of harassment, targets experience logistical and financial burdens in remediating the harassment, including getting new accounts and devices or even moving to a new home. 
Our participants experienced all of the above. 
When they were able to get adequate support, these issues were difficult but tolerable. Some individuals whose peers and institutions were less helpful have left the departments or lines of research which led to their harassment. 
Further, the capacity for this harassment to spill onto administrators, or for administrators to incur additional work in the course of assisting faculty going through harassment created additional professional consequences for participants.

\subsubsection{Personal Impact.} 
When harassment was personal and hateful, particularly when it aggravated existing marginalization -- i.e. misogynistic or racist harassment -- participants found it more difficult to overcome emotionally. 
Several participants cried in the course of their interview. 
Some felt they had been unable to avoid internalizing some of the hatred, others were overwhelmed by the depravity it displayed. 
Even in the face of this hatred, participants expressed that fear for their loved ones most limited them.
Participants who were parents indicated that concern for their children's safety kept them from doing work they otherwise might have; two participants who were not parents explicitly said they would not do the work they did if they were. E.g.:

\begin{displayquote}
``I don't think we could do this if we had kids. Honestly, I don't really know how you would do it if you had kids. Things get orders of magnitude more complicated.'' --- \textit{Studying Cyber Crime}
\end{displayquote}
While dealing with harassment was difficult, some participants felt it was something they had chosen for themselves in pursuing the research they did, and therefore, that they did not have a right to feel bad about it, or to expect sympathy from their colleagues. Participants discussed hiding the depth of their harassment from their colleagues and families, feeling it was their burden alone to bear. 
Three participants indicated that harassment spilled over onto departmental administrators, by way of voicemails, emails, or social media activity, and that they felt guilty about that. E.g.:

\begin{displayquote}
``I feel a bit guilty towards the person who runs the department Twitter account because I know that their @ will get flooded with nasty, racist, misogynistic shit if I tag them in posts. Even though it's not aimed at them, it's just not nice stuff to have around.'' --- \textit{Female, Tenured Professor}
\end{displayquote}

\subsubsection{Administrative Impact.}
\label{subsec:admin}
Many participants indicated that their institution did not have a concrete set of policies and procedures to handle harassment incidents. They felt overwhelmed by the harassment and the task of attempting to remediate it and felt let down or betrayed by their institution's lack of support.
Some participants found that the institution, or departmental leadership, failed to respond whatsoever, or were dismissive. 
In cases where participants felt their institution was making a good faith effort to respond,  several noted that the role they were asked to take in formulating policies going forward was unwelcome, and a burden unto itself. E.g.:

\begin{displayquote}
``You're the first one who's ever been harassed by weirdos to this level. Now you get to create the frickin' policy and get that through the provost's office and everything else.'' --- \textit{Computer Science Professor}
\end{displayquote}
All of the participants who described being asked to take on some of the work of formulating response policies and procedures, test new IT solutions, or give testimonials were female. 
This reflects recent findings~\cite{deo_2019} regarding women law faculty being given outsize amounts of `prestigious' committee work, which actually becomes a burden on their ability to accomplish scholarship.

When harassment involves or garners media attention, the institution's PR department may become involved.
Some of our participants viewed their PR department as a significant ally, prepping them to do adversarial TV interviews, or handling fire for them, while others felt the PR department was more concerned about covering for the university than protecting them, and that they would be asked to respond to media inquiries which were clearly in bad faith, creating an adversarial relationship between the professor and the communications department. Universities' communications departments are trained to handle bad press or refute misinterpretations of scientific results, but may not respond appropriately to media which is inherently adversarial. E.g.:

\begin{displayquote}
``I'm getting pushback from university communications - Infowars\footnote{InfoWars is an American far-right conspiracy theory and fake news website owned by Alex Jones \cite{infowars}}
 ran one on me saying [...] and university comms got upset and said, you know, `how do you respond to these very serious charges?' I was like: Did you just call Infowars very serious?'' --- \textit{Studying Alt-Right}
\end{displayquote}
The conflict of interest between the institution and the individual faculty member can be seen with respect to legal matters as well. 
When high-profile harassers use SLAPP\footnote{Strategic Lawsuit Against Public Participation are lawsuits filed by organizations with significant resources, not because they are winnable, but because fighting them is so burdensome it is easier to give in to demands.} suits, or threats of suits, as a silencing tactic, universities' legal departments do not always give full-throated defenses of academic freedom, instead seeking to protect the university from liability, \cite{FBSLAPP} often by requesting that the faculty member stop the work they are doing, or recommending they be terminated. E.g.:

\begin{displayquote}
``I really wanted some advice on libel, I was scared [of getting sued]. I tried to speak to the legal team [here]. It's a big university, you would expect the legal support to be quite good. I couldn't get them to give me even generic advice.'' --- \textit{Tenured Professor}
\end{displayquote}

\subsection{Constraints in Remediating Harassment}
Academics, journalists, activists, elected officials, and others who operate in the sphere of public consciousness are both more vulnerable to harassment and less able to mitigate it. 
The advent of social media has allowed a greater number of academics to be public intellectuals in ways that are beneficial to them, their departments, and arguably society, but also invite unprecedented harassment. The push towards public scholarship has meant researchers are increasingly professionally unable to remove themselves from the social media arenas in which they are likely to be harassed -- nor should they have to. This professional tension undermines much guidance on remediating harassment; one does not simply close their eyes, walk away from the screen~\cite{tylercreator}, and get off Twitter without significant career implications.
One participant explained that remediation advice largely involved removing oneself from the public sphere, which did not make professional sense:

\begin{displayquote}
``[Guides on what to do if you're harassed] don't really talk about academic freedom. They don't talk about the impact on your career from an academic point of view, a lot of them are just very clinical about how to take care of yourself and how to remove your name from services.'' --- \textit{Computer Scientist}
\end{displayquote}

\descr{Lack of support.} Participants explained that the support they received, or lack thereof, was fundamentally blind to this tension;
institutions and departments were at best ill-equipped to support faculty experiencing harassment, and often entirely dismissive of the issue, or resentful of an individual's choice to operate in the public sphere. E.g.:

\begin{displayquote}
``I had colleagues who basically told me `I'm sorry people are being mean to you on the Internet.' I was like, it's far more extensive than that. [...] There was just a real lack of understanding from my colleagues and the folks in administration on what was happening and what to do.'' --- \textit{English Professor}
\end{displayquote}
For other participants, it was even more discordant; their department appreciated the publicity, and therefore funding, that came with the public impact of their work, but was largely unwilling to provide resources to address the impact of harassment:

\begin{displayquote}
``I was quite upset and also quite angry. It just feels a bit sour thinking, you know, you're really happy to use my work for publicity. But when I was scared and targeted, you weren't giving me enough support then, were you?'' --- \textit{Female, Tenured Professor}
\end{displayquote}
Beyond the need to be present on social media in order to effect change, faculty have an obligation to be available to their students, and thus cannot remove all contact information from the Web.
Some participants were able to get support from IT at their institution: putting their email behind a login, allowing only those affiliated with the university to access it, or by adding filters to it.  

\descr{Action backfires.} When participants were unable to get support from within their department or institution, they were forced to take matters into their own hands, in some cases by seeking out counsel, support, or resources, and in others by attempting to address the harassment head-on. Participants found reporting content to Facebook or YouTube to be burdensome, and Twitter's responses to reports to be insufficient. Further, blocking people or reporting them provides feedback to them, and participants were appalled by the fact that harassers seemed to find that satisfying:

\begin{displayquote}
``What really, really didn't help was reporting stuff to Twitter, it just made me sadder. When you report a tweet, the person who's tweeted it gets notified. There's something really gross about seeing them literally brag about how they've been reported and it hasn't been upheld.'' --- \textit{Tenured Professor}
\end{displayquote}
Participants who fought back by publicly discussing their harassment or filing complaints found it could backfire; harassers that previously targeted them opportunistically may begin to view them as an adversary and escalate. Further, colleagues may view these actions as petty retaliation, and admonish them. A law professor being harassed by the legal community found that discussing it caused it to escalate:

\begin{displayquote}
``The second I talked back, it all got worse by an order of magnitude, now I'm just a person of interest. I have a reputation in academia as being aggressive and dramatic, and I'm really not.'' --- \textit{Law Professor}
\end{displayquote} 

\descr{Physical Security.}
Participants facing motivated harassers expressed frustration at how difficult it was to avoid their harassers being able to locate their home address. They discussed issues with public databases and tax records, lamented that they could not get additional protections, and discussed strategies including creating shell companies to own their houses, or buying them in cash to avoid mortgage records. E.g.:

\begin{displayquote}
``I bought a condo for the first time and a week later I already had hate mail sent to my place. Someone used public property purchase records to look me up.'' --- \textit{Tenured Professor} 
\end{displayquote}
Academics noted that harassers could easily locate their offices, often in buildings with little or no security.
Participants avoided spending time in their department -- the address public information, felt uneasy in classes -- anyone can view the course schedule, and asked for campus police to escort them after dark or attend their talks -- requests that were sometimes denied, or billed to their grants. Participants noted they were ``only researchers,'' but pointed out that one highly motivated or radicalized individual could pose a serious threat, and they felt vulnerable at work. E.g.:

\begin{displayquote}
``One time the FBI came to warn me that I was being targeted for a bombing. They hunted me down at my office, and they're like, ` you know, this is really not a good, safe office, anyone can just walk right in here.' Yeah, it's called a college campus. That's how it is. There's no locks.'' --- \textit{Studying Extremists}
\end{displayquote}

\subsection{Academic Freedom}

Participants viewed the risk of harassment as a threat to their academic freedom. Many indicated that past, ongoing, or even potential harassment had influenced aspects of their academic lives, including their choice of research topics, how they interact with students, other scholars, and the general public.
Here we discuss the effect of harassment on academic freedom in detail, looking at teaching, research, and public speaking, and how these issues interact with the tenure process.

\subsubsection{Teaching}
\label{subsec:undergrads}
Five participants discussed viewing undergraduate students as a threat, with `cancellation' as the primary vector. Harassment in the form of nebulous uproar from the student body may result in administrators siding with harassers. A conservative professor lamented that this was an emerging trend, particularly for people with his views:

\begin{displayquote}
``To be blunt, academic administrators too often are just craven, just cowardly; they feel the pressure coming from the mob, so they cancel, they fire, or they demote or take disciplinary action against the poor guy whose only crime is speaking his mind in a way that defies whatever the dominant orthodoxy is. Other academics don't hang around to support victims; rather, they scatter and go completely silent." --- \textit{Tenured Professor}
\end{displayquote}
Faculty with progressive politics worried about recordings being taken out of context, particularly in the era of Zoom lectures, and felt that was a serious impediment to their ability to exercise academic freedom in their classroom. 
Participants discussed students fabricating controversy to resolve grading disputes in addition to genuine ideological objections.
Two faculty indicated that right-wing media organizations were explicitly encouraging their students to catch them saying something inflammatory on camera, or even sending non-student reporters to their classes. E.g.:

\begin{displayquote}
``There was somebody who pretended to be a student and sat in on my classes to, like, write about me. They're interns for conservative think tanks or something.'' --- \textit{Studying Race Issues}
\end{displayquote}
Participants also noted that at institutions with an emphasis on teaching, or where student satisfaction, as measured by student feedback, is strongly connected to funding or tenure decisions, that this feedback could be an avenue of harassment, or could be leveraged by those with bad grades. Work has shown that in fields where teaching feedback is critical to tenure decisions, it further limits the career progress of women and people of color, who are more likely to receive negative feedback about any and everything, including their attitude and appearance~\cite{deo_2019}.
A participant described receiving censure as a result of a disingenuous student complaint:

\begin{displayquote}
``The institution is signed up to this student satisfaction idea because it's a market model. It only takes one person out of a hundred person class, and of course \textit{you} know the true story about why they're unsatisfied, but they're going to try and blame you. And simply, the institution won't stand by you, it will try to mitigate the PR damage, treating complaints without any real legitimacy as being dead serious. The [incident] I was involved in, I think the student should have just been kicked out of the university for his behavior. But the university tried to get him his money back and quieten him down. It's left me very untrusting of the organization moving forward.'' --- \textit{Tenured Professor}
\end{displayquote}

\subsubsection{Research}

Participants discussed changing their research agenda after a harrowing harassment experience, even when the work 
was some of the most impactful they had done. E.g.:

\begin{displayquote}
``The hate and harassment we got from 4chan made me decide to step back from future projects with that team. I liked that team very much, but I decided not to work on such sensitive projects anymore. It did have a positive effect on me professionally, that paper is one of my most cited.'' --- \textit{Studying 4chan} 
\end{displayquote}
Other participants said they had avoided contentious research in the past, waiting for the protection of tenure or until they had a level of career seniority which allowed them to do the work without seeking media attention. Participants indicated tiptoeing around a contentious topic, for example studying an ideology instead of directly studying the group that holds it.
One participant avoided a particular line of research until it became simply {\em ``too important}'':

\begin{displayquote}
``We knew those were sticky places to be doing research, so we were choosing to focus on other kinds of things. But in 2016, it became obvious that the newly elected President Trump was echoing the content from these websites. So I said, `OK, this is really important to go do, I'm going to do it anyways.' But I knew I was kicking a hornet's nest." --- \textit{Studying Conspiracy Theories} 
\end{displayquote}
When institutions are unable to adequately support researchers who are being harassed, those researchers may choose to pursue other lines of research altogether -- unable or unwilling to shoulder the burdens on their own -- or leave academia. 
Two participants indicated left a position at one institution to go to another because they were unable to get the support they needed to continue their work. Several participants said that if they had not been fortunate to find a position that offered them that support, they would have left the academy altogether. E.g.:

\begin{displayquote}
``Temperamentally, it would have been impossible for me to censor myself. I'm not claiming this as a virtue. Had I felt there's no way I could make it in academia without censoring myself, I just would have gone into another line of work. I could have gone into the practice of law or the insurance or used car business." --- \textit{Conservative Legal Scholar}
\end{displayquote}
The potential for harassment from outside the academy creates ethical concerns with respect to the involvement of graduate students, even when work may be otherwise IRB exempt.
In fields like Computer Science where large collaborations are common, faculty end up conducting large-scale projects alone which would otherwise have had large teams. 
One participant wrote their first contentious paper alone; when they brought PhD students onto a follow-up paper, they continued to list themself as first author in an effort to divert harassment, though it would be customary to list the students.
One professor had a grant application denied because of the risk to students: 

\begin{displayquote}
``[The grant applications] were both rejected, unfortunately. One of the main comments the reviewers had was `you did not provide a plan for keeping grad students safe.' And so I think if even an NSF panel can realize this is a problem, we've got to solve this.'' --- \textit{Studying Far-Right}
\end{displayquote}
Failing to protect researchers from harassment puts them in a position where they may feel the need to choose their safety and sanity over their research. For researchers studying issues of race and gender, allowing the propensity for harassment to keep the work from being done is not only a concern of academic freedom but also deleterious to the interests of marginalized groups. E.g.:

\begin{displayquote}
``There's also things I don't write about now because I know that it's just going to set off a shitstorm. A lot of my [research] ideas are about the thing that's happened to me, but instead of writing about the way women are treated online, I've chosen to write about other things. --- \textit{Female, Tenured Professor}
\end{displayquote}

\subsubsection{Speaking Publicly}
Participants said media attention, particularly social media, was the primary instigator of harassment. 
Many felt they could avoid most of their harassment by staying off social media, and some had done so intermittently and seen it to be effective.
Often in the same breath, these individuals acknowledged that this created professional issues for junior faculty, and amounted to self-censorship regardless.
Participants who felt an obligation to maintain a public dialogue about their research said they were willing to take the fire, but might suggest their students keep off Twitter.
Participants also discussed being more cautious about advertising speaking engagements, and strictly limiting attendance information for online presentations to avoid harassment, thus limiting the potential reach of their talks. E.g.:

\begin{displayquote}
``I already had a reputation. I had done a lot of work. I could choose not to have [the talks] recorded, whereas if you're a younger scholar, it's good to have that public persona and record. [...] This research is important. Part of the problem is that people who are doing research in this area feel like they have to be pretty public about it and comment on it. And that's also just our milieu today.'' --- \textit{Anthropologist}
\end{displayquote}
The dampening and silencing impacts of harassment also interplay with existing marginalization: women and people of color are more likely to experience harassment than their counterparts~\cite{espinoza2018cyberbullying,zych2015scientific,gardiner2018sa}. Several participants addressed this, directly and indirectly. A professor told us he avoided projects in gendered spaces because it would yield too much gendered harassment for the women in his lab. A woman told us that on a paper with several male co-authors, she was the only one who had received any serious harassment. Re-aggravating marginalization can be a tactic, as one man that had the ``Alt-right gaze''~\cite{doi:10.1177/2056305118768302} turned on him explained how things played out in his diverse research group:

\begin{displayquote}
``I'm a person of color amongst a bunch of white researchers, so that was a strong distinguishing factor for me, and it played a role [in the harassment]. Another person was portrayed as a Jew -- and this person is not Jewish -- and they brought in some of the negative connotations against Jews. There was a lady among us who I know got hateful comments on the basis of her gender as well. Our team was harassed differently, we received harassment of various types, depending mostly on our [demographics]. And also, of course, how visible we were. Like for some people, things went really crazy like that. For some other people, not so much.'' --- \textit{Computer Science Professor}
\end{displayquote}
If marginalized scholars are not able to receive adequate support, it is likely that their white and/or male colleagues will be the ones who are able to continue these lines of research, be featured in media, or have a public presence. 
Thus, individuals already facing structural barriers to success in academia may be weighed down further: it is precisely the research that creates controversy that creates citations, and others can do it to a significantly less personal detriment. One man noted this in discussing a prominent woman of color in his space and the differences in their experiences:

\begin{displayquote}
``Being a predominantly white male, it's measurably less significant than for people who write about the same topic, who are more obviously from a minority group or women.'' --- \textit{White-Passing Man}
\end{displayquote}
Future work could specifically investigate the roles of race, gender, and other aspects of marginalization as they relate to harassment of researchers, in the manifestation, propensity for victimization, and severity of harms.

\subsubsection{Tenure}

Tenure is, generally speaking, \emph{the} major career milestone for most academics in many educational/research systems such as that in the US. %
In simplistic terms, tenure is often portrayed as the point where professors become ``unfireable.''
With more nuance, tenure affords certain (at least perceived) privileges and protections that enable true academic freedom.
That is, once tenured, not only are we free to pursue whatever intellectual endeavors we find interesting but are also protected from the negative effects of work that might be considered politically undesirable.
Attaining tenure is, of course, no easy task and junior faculty are expected to prove themselves worthy of tenure across a variety of axes, 
such as scholarship, %
teaching, and service. %

In addition to these relatively clearly articulated expectations, there are often unwritten rules.
Junior faculty may be wary that ``rocking the boat,'' and attracting negative attention from senior administration could impede their tenure case.
Participants noted that these unwritten rules with respect to political work created additional stress in an already stressful process:

\begin{displayquote}
``I wish that there had been more explicit things said to me about like, what lines I can and cannot cross as someone who was untenured because I think I lived in a space of a lot of fear of not being tenured simply because of my political activity.'' --- \textit{Tenured Professor}
\end{displayquote}
There was a sense that not only the type of work, but specific findings that might run afoul of the political norms of the academic community and general public can take a heavy toll on early career academics.
For example, the risk of backlash from the academy itself can dissuade researchers from pursuing academic careers in the first place, and has chilling effects on the ability of junior faculty to disseminate findings that could result in career-ending, politically motivated campaigns:

\begin{displayquote}
``Doctoral students are absolutely terrified. They see the terrible things that happen to people who have been canceled, because of their views and because a certain dogma has taken hold in the academy. [...] It makes it very hard for people who dissent from it, especially younger people, assistant professors who don't yet have tenure, to have the boldness to speak.'' --- \textit{Conservative Tenured Professor} 
\end{displayquote}
Tenure track positions are notoriously competitive, with orders of magnitude more applicants than positions available.
Generally speaking, tenure track candidates are expected to have impactful research agendas.
Yet, participants also noted that even when their work was well known and highly regarded, tenure track positions remained elusive in light of controversy:

\begin{displayquote}
``I'm permanent but nontenured. In Europe, I'm considered one of the key figures, and I can't get a tenure track job in this country.'' --- \textit{Queer Art Professor}
\end{displayquote}
Finally, it is crucial to note that the concept of tenure differs between geopolitical regions.
While tenured professors in the US system are essentially ``unfireable,'' this is not the same in every higher education system.
Although tenure comes with certain privileges in all systems that have it, participants noted that even after being tenured there were still risks posed by their work:

\begin{displayquote}
``It's a legitimate concern for somebody that doesn't have a UK passport, who, although I'm tenured, it's not the same as America. They still can get rid of you.'' --- \textit{Professor in the UK}
\end{displayquote}

\section{Recommendations}

In this section, we provide recommendations for individuals experiencing harassment, and for departments and institutions to support their faculty and students should research attract harassers. 
Further, we acknowledge that conflicts of interest exist between institutions and their faculties' research agendas. We suggest that a body with broader leverage take on the role of creating guidelines for institutions, and of providing advocacy support to faculty when these conflicts arise.

\subsection{Cultural Issues with How Academia Deals with Harassment}

One theme that pervades our recommendations, whether at the level of a PI, a department, an institution, or the academy as a whole, is the need for a cultural shift away from attitudes that exacerbate the problems we have outlined in this work. 
Multiple participants expressed that they are simultaneously lauded by their institutions for their cultural influence, derived largely from their public presence, and derided for that same public presence when the harassment it invites becomes burdensome for the organization. 

Ultimately, the goal of research is impact. We argue that media work, including social media, is a part of modern scholarship and an important part of science communication. %
Whereas some faculty may not choose to be active in the public sphere, those who do should be given the support they need to effect a positive impact on society, research, policy, and potentially departmental funding.
Such support can make a difference in how academics cope with harassment.
We noted that participants who had experienced egregious harassment, but had support from their institution, communicated more optimism and continued in lines of contentious research:

\begin{displayquote}
``[My institution] has been super supportive, very clear that they support my freedom of speech, my right to do whatever I want on my own time. And so I think the fact that I know that I'm supported and that I'm not going to lose my job allows me to continue to be as public as I am and in some ways kind of flaunt it.'' --- \textit{Female, Tenured Professor}
\end{displayquote}
Similarly, many individuals expressed that empathy from colleagues, or the lack thereof, was a critical factor in how they were able to respond to harassment. There is a need for greater awareness of the temerity of harassers---they are not simply ``people being mean to you on Twitter''---and a need for researchers to support each other and rally to each others' defense. %
Marginalized scholars being inundated with hatred from anonymous trolls or public figures should expect peers with more privilege to rally in their defense; similarly, conservative scholars being mobbed by students should expect their liberal counterparts to publicly reject harassment as a way of silencing scholars.  

\begin{displayquote}
``Freedom of speech is important for everyone, but especially for scholars because of their vocation to be truth-seekers, to be poking and prodding and questioning orthodoxies. So when our fellow academics are unjustly attacked and threatened for challenging dominant views and speaking their minds, we've got to start rallying to each others' defenses.'' --- \textit{Conservative Professor}
\end{displayquote}
One professor expressed feeling angry and betrayed after colleagues said nothing in the wake of a harassment campaign against her, or even that those that did, did so in private:

\begin{displayquote}
``I'm still so just enraged at the more senior people in the academy who saw what was going on and didn't support me when I was just, you know, getting rape threats... I actually feel like most people in academia agreed with me. But here's the thing, I'm mad at them, too, because, like, they didn't want to be harassed. Like they didn't say anything publicly, they were quiet.''  --- \textit{Female, Tenured Professor}
\end{displayquote}
She has since seen similar attacks happen to junior colleagues, and despite the personal risk, insists on speaking out publicly to put an end to it. Others discussed going to bat for their students.

One participant's work was being patently censored after they became a target of a Republican politician in an election year. 
Instead of backing down or filing suit, the professor solicited support from media outlets, cultural institutions and advocacy organizations---to great success:

\begin{displayquote}
``I wanted to fight this in the court of public opinion, and that's what I did. I set up a website, I started traveling across the country delivering talks, and I was able to mobilize their censorship against them. And successfully. [The politician] lost his re-election bid. --- \textit{Queer Art Professor}
\end{displayquote}

\subsection{Institutions and Departments}

A relatively small fraction of researchers will ever experience harassment as a result of their work, and fewer still will experience harassment which rises to the level described by our participants. Yet, it is bound to happen. We believe that particularly at large institutions, it is appropriate to have an officer who is equipped to handle harassment when it arises. Such a role requires broad adoptions of guidelines and policies that do not yet exist, despite efforts to articulate recommendations~\cite{DSguide}.

Training in recognizing and remediating harassment could be integrated into existing standard training protocols for people engaged in research. There is also a need for interdisciplinary working groups that understand the diverse set of motivations and harassment tactics. 
PR departments need guidance on handling adversarial media and an understanding that some quasi-media organizations are adversarial. 
Policies should be clear about professors' legal protections and resources provided to defend against threats, including libel suits and other forms of strategic lawsuits against public participation (SLAPP) from major corporations, political parties, and other well-established entities. 

Mental health services ought to be made readily available to anyone engaged in research that has toxic imagery or content, or who experiences harassment as a result of their work. Departments should anticipate potentially reduced productivity on projects from which PIs and students need to step back to recover from harassment-related harms. Crucially, the outcome of such institutional guidelines and policymaking is that researchers should feel they have allies in their institutions.

\subsection{Individuals}

PIs should consider what hornet's nests they may be kicking, understand what parties may take issue with their research, and how those parties tend to react. Some scholars are well attuned. For example, crime researchers may expect harassment from their subjects, and race researchers may expect intimidation by white supremacist groups.
 Harassers have varying motivations and modi operandi; we hope that our findings of who commits harassment of researchers and what their motivations are are helpful in moving towards a more universal awareness. 

Individuals should take standard digital hygiene and security steps, such as enabling two-factor authentication on accounts that attackers may attempt to breach, and may consider measures such as scrubbing old posts, although it amounts to self-censorship.
There are Twitter settings that a number of our participants discussed being helpful, though some of these have censorship implications, so we discuss them without inherently suggesting them. 
Some people blocked tweets with certain keywords (e.g. slurs), some individuals only allowed replies from followers, or sometimes, only from those whom they follow. 

One participant found a lot of hate stemmed from articles about them being spread on Facebook, and indicated they have since gotten access to software that makes it much easier for them to report these things, as it is otherwise quite burdensome:
\begin{displayquote}
``I didn't have that [software] back then, so a lot of those people just were able to continue harassing other people because no one had reported it. ... It's just a hassle to report those things because on Facebook it takes 10 clicks to report anything.'' --- \textit{Computer Science Professor}
\end{displayquote}
PIs should take reasonable steps to protect their students and collaborators but should avoid unduly limiting those individuals. 
Good practices include having a frank discussion with students about the potential risks of their research and being open to discussing their harassment experiences~\cite{DSguide}. 
Additionally, faculty employed many strategies in an effort to protect their students and other vulnerable colleagues from the ramifications of divisive work including: excluding others in favor of working alone, foregoing projects altogether, altering authorship orders to draw fire, siloing traumatizing pieces of the work, and encouraging students to refrain from social media and publicity campaigns.
Rather than taking paternalistic actions to protect more vulnerable colleagues, we recommend engaging in frank discussions of risk so that students or marginalized colleagues can make informed decisions about their participation.

Finally, many participants reported that the most effective strategy was simply not engaging, not reading comments, or blocking people who were aggressive. 
Again, there are censorship implications.
Many participants also extolled the virtues of a thick skin and the attitude of ``not taking it too seriously.'' They were overwhelmingly male.

\subsection{Advocacy and Resources Outside One's Institution}

When the interests of researchers and their institution are at odds, there may be a need for a third-party advocate to support the researcher with legal and other resources. For example, if research is met with backlash from donors, major corporations, politicians, and other groups with significant cultural heft. Organizations like the American Civil Liberties Union (ACLU), Foundation for Individual Rights in Education (FIRE), %
Union of Concerned Scientists, and American Association of University Professors (AAUP) are candidates, or at least models for taking on such a role.

When there are no direct conflicts of interest, there is still a need for some organizations to take on the work of creating guidelines and standards of practice which institutions, departments, and PIs can follow. 
Several participants noted being the first at their institution to receive harassment, and subsequently becoming in charge of the non-trivial project of developing such policies and procedures, as discussed in ~\ref{subsec:admin}. 
One participant reported that researcher harassment has been used as grounds for rejecting grant proposals; 
it would be useful for funding organizations to both have access to and provide recommendations on threat mitigation.

Finally, the creation of a peer-to-peer network, such as Heartmob~\cite{Heartmob}, specifically for researchers, would be helpful here as well, as many individuals expressed that kindness from colleagues was a primary mechanism of support for them. 

\subsection{Future Work}
All of these recommendations have cultural and institutional aspects, but there is still work needed to fully understand the scope of this problem and its harms, as well as the viability and efficacy of potential solutions. In this work we have made an effort to include diverse voices and explore the ways in which internet-facilitated harassment interacts with and exacerbates existing marginalization in society and academia, but further work is needed to understand these relationships specifically, including the ways in which these experiences are gendered and racialized. 

Heartmob has been studied as an effective tool for providing support to victims of harassment, but its beneficiaries are often using the service and being harassed in online spaces with pseudonymous usernames, such as Tumblr. The creation and study of a researcher-specific peer network could help examine support structures for those who are in the public eye with their real names and reputations on the line. 

This work has focused on the experiences of faculty and similar individuals, but in the interest of fairness, work that focuses on administrations and administrators is merited. While we may have significant criticism here of the responses of institutions, institutions and administrations are composed of individuals, and it would be worthwhile to understand how they view and interact with these issues, and the constraints they face that keep them from reacting in ways which we may find more beneficial to faculty. Finally, deeper research on effective policies needs to be done, in the interest of creating a set of standards that can be adopted by institutions and grant funders, without placing the burden of remediation and mitigation on those experiencing harassment.

\section{Conclusion}
Through interviews with 17 individuals in academia, journalism, and advocacy, we developed a framework for understanding the types of harassment that befall researchers as a result of their work and the motivations of their harassers. 
We found that there are three overarching categories of harasser motivations: self-preservation, ideology, and performance. 

Overall, by understanding the different parties who engage in harassment, and why, we can better anticipate what work may incur harassment, and the form that may take -- a critical step towards protecting researchers and research, as different types of harassment have different consequences and necessitate different remediation strategies.

We examined what has been effective for researchers in preventing or remediating harassment, and more so, what has not; many of the known best practices for mitigating and remediating harassment are at odds with the professional needs of researchers and others living in the public eye, particularly for junior academics. 
Further, we found that researchers are hamstrung by the lack of support available to them in the face of attacks from both outside and inside the academy. 
Faced with the choice to stop doing divisive work, stop being public about it, or trudge through the vitriol,  researchers choose all three; the first two having significant ramifications for academic freedom, and the latter being more accessible to those with more seniority and structural privilege.

While harassment cannot be stopped, its detrimental effects can be limited, and therefore its impact of public scholarship managed. 
We offered a set of recommendations for managing safe, controversial research, and for protecting researchers as well as academic freedom, as well as a plea for others to take up the future study of the nuances of this issue, and to develop formal guidelines that can be adopted throughout the academy.

\descr{Acknowledgments.} The authors wish to thank the participants as well as the people who helped recruiting them.
This research was supported by the National Science Foundation (grant numbers 2016061, 1942610, 2114407, and 2114411) and REPHRAIN: The UK's National Research Centre on Privacy, Harm Reduction, and Adversarial Influence Online (UKRI grant: EP/V011189/1).
\vspace*{-0.4cm}

\small
\bibliographystyle{abbrv}

\appendix

\begin{table*}[t]
\begin{center}
\renewcommand{\arraystretch}{0.86}
\setlength{\tabcolsep}{3pt}
\resizebox{0.99\textwidth}{!}{
\begin{tabular}{@{}l@{}rl@{}lr@{}}
\toprule
\textbf{Super Category} & \textbf{Sub Category}            & \textbf{Details}                                 & \textbf{Acting Party}                                             & \textbf{\#} \\ \midrule
academic                & censorship                       & publications censored                            & artist's foundation                                               & 1           \\
                        &                                  & barred from academic venue                       & artist's foundation                                               & 1           \\
                        &                                  & kicked out of a talk                             & artist's foundation                                               & 1           \\
                        &                                  & Unauthorized editing of video lectures           & host of lecture                                                   & 1           \\
                        & interference with career         & PhD program defunded                             & dean                                                              & 1           \\
                        &                                  & campaign to have work censored                   & GOP                                                               & 1           \\
                        &                                  & Calls for firing                                 & right wing, left wing                                             & 4           \\
                        &                                  & malicious survey responses                       & -                                                                 & 1           \\
                        &                                  & visa cancelled                                   & foreign government                                                & 1           \\
                        &                                  & threat to journal after publishing work          & conspiracy theorists                                              & 1           \\
                        &                                  & manipulation of research                         & white supremacists, conspiracy theorists                         & 2           \\
                        &                                  & threat of lawsuit                                & far right leaders and groups                                      & 3           \\ \hline
cyber attack            & account compromise               & spear phishing                                   & cyber criminals                                                   & 1           \\
                        &                                  & social engineering of customer service           & cyber criminals                                                   & 1           \\
                        &                                  & attempted account compromise                     & cyber criminals, conspiracy theorists, far right                  & 3           \\
                        & doxing                           & doxing of friends, family                        & cyber criminals                                                   & 1           \\
                        & doxing                           & doxing                                           & cyber criminals, hate groups, 4chan                               & 3           \\
                        & other                            & DDoS attack                                      & cyber criminals                                                   & 1           \\ \hline
in-person               & demonstration                    & picketing on campus                              & neo-nazi groups                                                   & 1           \\
                        & intimidation                     & stalking                                         & -                                                                 & 2           \\
                        &                                  & unwanted police presence at meetings             & police                                                            & 1           \\
                        &                                  & non-students attending lectures                  & right wing, conspiracy theorists                                  & 2           \\
                        &                                  & confederate flag planted in yard                 & right wing                                                        & 1           \\
                        &                                  & breaking and entering                            & -                                                                 & 1           \\
                        & verbal                           & comments from colleagues                         & colleagues                                                        & 1           \\
                        &                                  & in-class disruption                              & students                                                          & 1           \\
                        &                                  & combative Q\&A after a talk                      & white supremacist                                                 & 1           \\
                        &                                  & came to department and spoke to supervisor       & conspiracy theorists                                              & 2           \\
                        & weaponization of  & SWAT raid                                        & cyber criminals                                                   & 1           \\
                        & \hspace*{0.15cm} law enforcement                                 & drugs delivered to house                         & cyber criminals                                                   & 1           \\ \hline
media                   & high-profile blogs, newsletters  & antagonistic stories                             & far right, campus republicans                                     & 6           \\
                        & news sites                       & antagonistic stories                             & state-sponsored media, far left, far right                        & 5           \\
                        & television                       & antagonistic stories                             & Fox News                                                          & 1           \\ \hline
online hate             & 4chan \& similar                 & threads used to organize harassment elsewhere    & 4chan                                                             & 2           \\
                        &                                  & threads on 4chan full of hate                    & 4chan                                                             & 3           \\
                        & email                            & signed up for hundreds of listservs              & -                                                                 & 1           \\
                        &                                  & hate mail                                        & -                                                                 & 10          \\
                        &                                  & emailing institution, department, etc            & -                                                                 & 3           \\
                        & Facebook                         & viral posts about you                            & far right leaders                                                 & 1           \\
                        &                                  & Facebook groups used to organize harassment      & -                                                                 & 1           \\
                        &                                  & fake profiles sending messages                   & -                                                                 & 1           \\
                        &                                  & comments on Facebook page {[}of org, not self{]} & white supremacist group                                           & 1           \\
                        & other apps                       & Telegram channels used to organize harassment    & -                                                                 & 1           \\
                        & other websites                   & Wikipedia page edited                            & -                                                                 & 1           \\
                        &                                  & hit pieces on personal blogs                     & colleagues                                                        & 1           \\
                        &                                  & abuse on small, topic-specific blogs and forums  & far right, legal field                                            & 2           \\
                        &                                  & abusive comments (authorized/authored articles)  & -                                                                 & 3           \\
                        & Twitter                          & accounts dedicated entirely to harassing you     & -                                                                 & 2           \\
                        &                                  & DMs                                              & cyber criminals, far right leaders                                & 2           \\
                        &                                  & hateful @s                                       & -                                                                 & 9           \\
                        &                                  & viral posts about you                            & journalists, far right leaders, far left leaders                  & 4           \\
                        & YouTube                          & videos of talks uploaded without consent         & attendee                                                          & 2           \\
                        &                                  & hateful videos about you                         & vlogger                                                           & 2           \\
                        &                                  & comments on videos of talks                      & -                                                                 & 4           \\
                        & Zoom bombing                     & screen sharing                                   & -                                                                 & 1           \\
                        &                                  & drawing on screen                                & -                                                                 & 3           \\
                        &                                  & chat box                                         & -                                                                 & 2           \\
                        &                                  & speaking                                         & -                                                                 & 2           \\
                        &                                  & profile or background picture                    & -                                                                 & 2           \\ \hline
phone calls             & calls to others                  & calling collaborators                            & conspiracy theorists                                              & 1           \\
                        &                                  & calling administrators                           & far right                                                         & 2           \\
                        & calls to researcher              & verbal abuse                                     & colleagues/peers                                                  & 2           \\
                        &                                  & death threats                                    & activist                                                          & 1           \\ \hline
reputational            & looking for gotchas              & proposals from prospective PhD students          & trolls                                                            & 1           \\
                        &                                  & invitations to adversarial fora                  & right wing                                                        & 2           \\
                        &                                  & unauthorized attendees to classes/lectures       & right wing organizations, conspiracy theorists                    & 2           \\
                        &                                  & unauthorized recording of class or talks         & right wing organizations                                          & 3           \\
                        & slander campaigns                & with goal of harming reputation within field     & media, colleagues, students, conspiracy theorists                 & 6           \\
                        &                                  & with goal of harming public image                & criminals, media (left and right), students, conspiracy theorists & 11          \\ \hline
snail mail              & dangerous objects                & drugs delivered to house                         & cyber criminals                                                   & 1           \\
                        &                                  & unidentified white powder (office)               & nazi group                                                        & 1           \\
                        & letters                          & unsolicited magazines (office)                   & far right                                                         & 1           \\
                        &                                  & forwarded junk mail (office)                     & far right                                                         & 1           \\
                        &                                  & hate mail sent to office                         & nazi groups, far right, far left, conspiracy theorists            & 6           \\
                        &                                  & hate mail sent to home                           & hate groups, far right                                            & 3           \\
                        & packages                         & gifts sent to partner                            & cyber criminals                                                   & 1           \\
                        &                                  & sent to office                                        & inmates                                                           & 1           \\
                        &                                  & drop shipped from Amazon                         & -                                                                 & 1           \\ \hline
threats                 & blackmail                        & intimate imagery {[}including shopped{]}         & 4chan                                                             & 2           \\
                        & physical violence                & bomb threat {[}work{]}                           & nazi group                                                        & 1           \\
                        &                                  & bomb threat {[}home{]}                           & hate group                                                        & 1           \\
                        &                                  & credible death threats                           & cyber criminals, nazi groups, far left                            & 3           \\ \bottomrule
\end{tabular}}
\caption{Vectors of Harassment.}
\label{table:huge}
\end{center}
\end{table*}

\end{document}